\documentclass[11pt]{article}

\usepackage[utf8]{inputenc}
\usepackage{amsmath,amssymb,amsfonts} 
\usepackage[UKenglish]{babel} 
\usepackage[T1]{fontenc}
\usepackage{setspace}
\usepackage{xcolor}
\usepackage{graphicx}
\usepackage{lmodern}
\usepackage{booktabs}
\usepackage{tabularx}
\usepackage{caption}
\usepackage{hyperref}
\usepackage[T1]{fontenc}
\usepackage{ae,aecompl}
\usepackage{subfig}
\usepackage{fullpage}
\usepackage[toc,page]{appendix}
\usepackage{epstopdf}
\usepackage{import}

\graphicspath{ {Allplot/}{figures/} }

\newcommand{\md}{\mathrm{d}}
\newcommand{\pd}[2]{\frac{\partial #1}{\partial #2}}

\newcommand{\suml}[2]{\sum\limits_{#1}^{#2}}
\newcommand{\intl}[2]{\int\limits_{#1}^{#2}}

\newcommand{\vy}{ v_{y}}

\newcommand{\C}{ \mathcal{C}}
\newcommand{\p}{\preceq}
\newtheorem{claim}{Claim}
\newtheorem{conjecture}{Conjecture}
 
\newtheorem{definition}{Definition}

\newcommand{\G}[1]{\Gamma\left(#1 \right)}
\newcommand{\mFm}[4]{\, _{#1}F_{#1}\left( \genfrac{}{}{0pt}{}{#2}{ #3  }  \bigg| #4 \right) }
\newcommand{\Poch}[2]{\left( #1 \right)_{#2}}
\newcommand{\vol}{\mathrm{Vol}}
\newcommand{\re}{\mathbb R}
\newcommand{\pFq}[5]{\, _{#1}F_{#2}\left( \genfrac{}{}{0pt}{}{#3}{ #4  }  \bigg| #5 \right) }
\newcommand{\NmC}{N_m(C')}
\newcommand{\NmCC}{N_m(C)}
\newcommand{\Nmd}{\langle N_m^d \rangle}
\newcommand{\Nmdv}[2]{\langle N_{#1}^{#2} \rangle}

\newcommand{\Nzerod}{\langle N_0^d \rangle}

\newcommand{\pprec}{\prec\prec} 
\newcommand{\curv}{\zeta}
\newcommand{\tC}{\widetilde C} 
\newcommand{\NmtC}{N_m(\tC)}

\newcommand{\Sc}{\mathcal S}
\newcommand{\nno}{\nonumber}
\newcommand{\Cm}{\langle C_m\rangle}

\title{Towards a Definition of Locality \\ in a Manifoldlike Causal Set} 
\author{Lisa Glaser$^{a}$ and Sumati Surya$^{b}$\\ 
$^a$ Niels Bohr Institute, Copenhagen, Denmark \\
$^b$ Raman Research Institute, Bangalore, India}
\begin{document}

\maketitle

\begin{abstract}
  It is a common misconception that spacetime discreteness necessarily implies a violation of local
  Lorentz invariance.  In fact, in the causal set approach to quantum gravity, Lorentz invariance
  follows from the specific implementation of the discreteness hypothesis. However, this
  comes at the cost of locality. In particular, it is difficult to define a ``local'' region
  in a manifoldlike causal set, i.e., one that corresponds to an approximately flat spacetime
    region.  Following up on suggestions from previous work, we bridge this lacuna by proposing a
  definition of locality based on the abundance of $m$-element order-intervals as a function of $m$
  in a causal set.  We obtain analytic expressions for the expectation value of this function
  for an ensemble of causal set that faithfully embeds into an Alexandrov interval in 
  $d$-dimensional Minkowski spacetime and use it to define local regions in a manifoldlike causal
  set.  We use this to argue that evidence of local regions is a necessary condition for
  manifoldlikeness in a causal set.  This in addition provides a new continuum dimension estimator.  We
  perform extensive simulations which support our claims.
\end{abstract}


\section{Introduction}
Causal set theory is a candidate for quantum gravity where the spacetime continuum is replaced by a
discrete substructure which is a locally finite partially ordered set \cite{cst,valdivia,
  Dowker:2005tz, Surya:2011yh}.  It is often assumed that Lorentz violation is an inevitable consequence of
spacetime discreteness. This is explicitly false in a causal set discretisation of a spacetime -- on
the contrary, as shown in \cite{bomhensor}, the causal set hypothesis instead {\it implies} Lorentz
invariance.  However, this comes at the cost of locality.  In a causal set, the nearest neighbours
of an element are the links or irreducible relations. For example, in  infinite causal set that is
approximated by Minkowski spacetime, every element has an infinite number of nearest neighbours,
both to the past and to the future. The resulting graph is therefore of infinite valency, in stark
contrast to other types  of spacetime discreteness in which the graphs are of finite valency.
Indeed, it is this very feature of a causal set which captures the essence of Lorentz-invariant
discreteness, since there are non-compact invariant hyperbolae associated with every Lorentz boost
about a spacetime event in Minkowski spacetime.
This feature of causal set discretisation in turn is due to  the requirement of a uniform
distribution which preserves the number to volume correspondence,  crucial to the recovery
of the Lorentzian spacetime geometry in the continuum approximation \cite{bomhensor}.

While Lorentz invariance is a great asset to causal set theory, the resulting non-locality of the
causal set graph impedes a straightforward reconstruction of continuum information from the discrete
substructure. Unlike a simplicial decomposition, for example, where the discrete scalar curvature
has a simple {\it local} geometric interpretation, there is no analogous local construction in a
causal set. Indeed, it is only recently that a causal set expression for scalar curvature and hence
a causal set action has been found in arbitrary dimensions
\cite{Benincasa:2010ac,Dowker:2013vba}. Nevertheless, despite the difficulty in recovering local
information from a causal set, over the years substantial progress has been made in understanding
how topology and geometry emerges from a causal set, sometimes with the aid of fairly ingenious
order-theoretic constructions. This includes the reconstruction of spacetime dimension, time-like
distance, space-like distance and spatial homology, for causal sets that are approximated by
continuum spacetimes \cite{allkinematics}.

Importantly, in many of these reconstructions the causal set is assumed to be approximated by a
region of curved spacetime which is small compared to the scale of flatness. In the continuum such a
region has a natural interpretation of being ``local''  or approximately flat.  From the
continuum perspective small, or local neighbourhoods are essential to several geometric
constructions and are key to the conception of a manifold. However, until now there has been
  no purely order theoretic characterisation of such local neighbourhoods in a causal set.  It is
therefore an important step to be able to define local regions in a causal set and  hence provide an
appropriate context for some of the reconstruction results. 

\begin{figure}[ht]
\vspace{1cm} 
\centering \resizebox{2.5in}{!}{\includegraphics{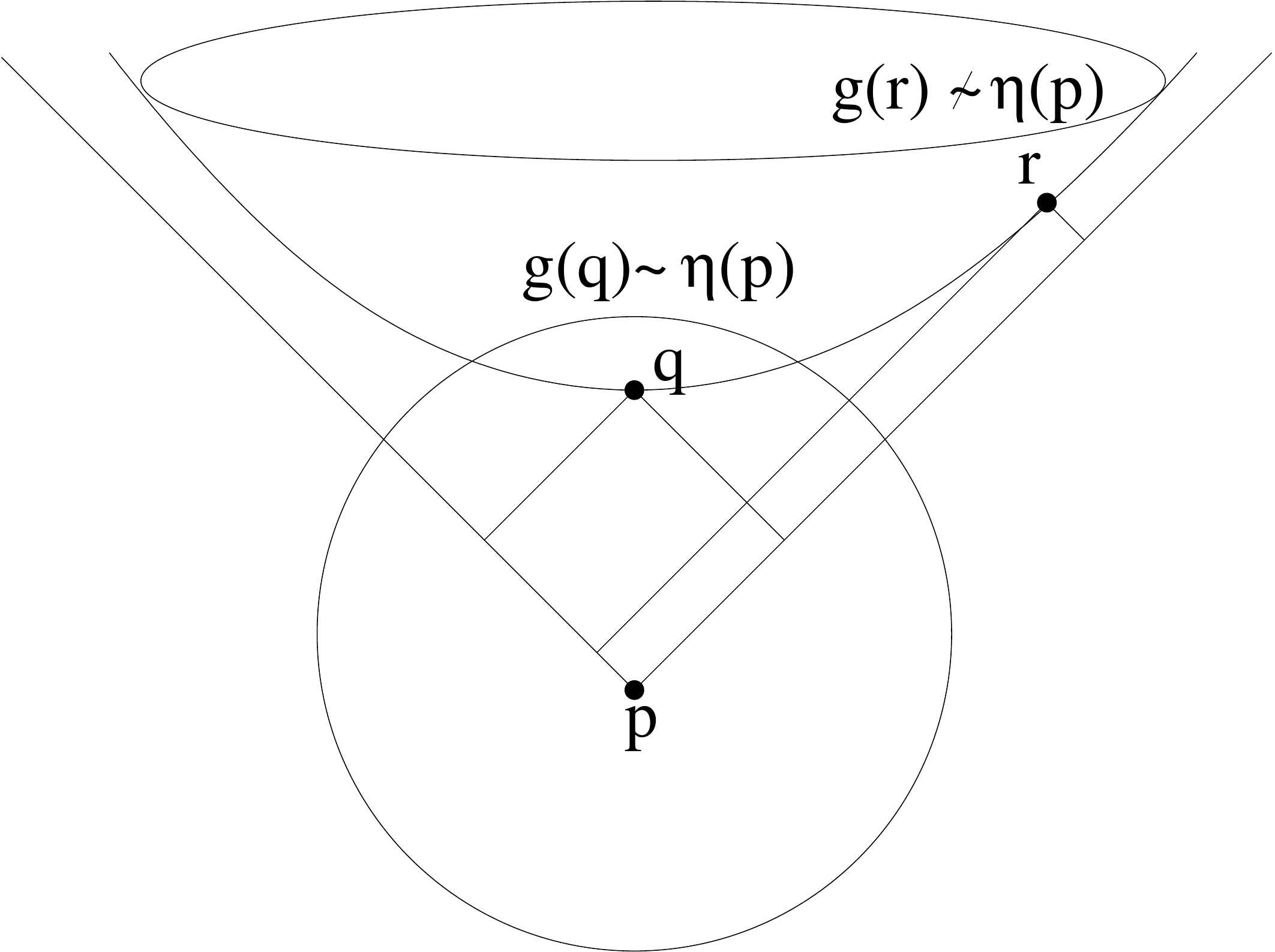}}
\vspace{0.5cm}
\caption{\small The two Alexandrov intervals $I[p,q]$ and $I[p,r]$ have the same volume, but the
  former lies in an approximately flat neighbourhood of $p$,  while the latter does not.}
\label{fig:alex}
\end{figure}

Our prescription for locality uses a well known order theoretic definition of a spacetime region,
namely an Alexandrov interval $I[x,y]:=\{z| x\pprec z \pprec y \} $, where $\pprec$ is the
chronological relation. In the continuum $I[x,y]$ is characterised both by the time-like distance
$\tau(x,y)$ from $x$ to $y$, as well as its volume $\vol(I[x,y])$.  Unlike open ball
neighbourhoods in a Riemannian manifold, however, even arbitrarily small choices of $\tau(x,y)$ or
$\vol(I[x,y])$ do not correspond to a region in which  the scale of flatness is large as
illustrated in Fig. \ref{fig:alex}. In the continuum, $\tau(x,y)$ or $\vol(I[x,y])$ are the only Lorentz
  invariant quantities that characterise $I[x,y]$. However, it is clear  that the corresponding  discrete geometry, i.e., 
  a causal set which faithfully embeds into $I[x,y]$,  should contain more detailed 
  geometric information.   

The main proposal of our work is that there is indeed an order theoretic  characterisation of locality.
The motivation for this arises from the work of Benincasa and Dowker
  \cite{Benincasa:2010ac,Dowker:2013vba} where it was found that the scalar curvature of an element
  $e$ in a causal set $C$ can be obtained from knowing the abundances $N_m$ of order-intervals of
  size $m$ that lie to the  past of $e$. Here, $N_0$ is the abundance of  $0$
order-intervals, i.e., the number of irreducible relations or links in $C$, $N_1$ the abundance
   of $1$-element order intervals, or irreducible 3-chains etc.For a generic  non-locality scale  the  discrete
    Einstein-Hilbert action is constructed from all possible $N_m$, but   when the
    non-locality scale is taken to be the Planck scale, the action simplifies considerably. For example,
    the 2d causal set action
takes the elegant form
\begin{equation}
\frac{1}{\hbar} S_{2D} = N- 2  N_0 +  4 N_1 - 2 N_2   
\end{equation}
which only involves the abundances of intervals of volume $0,1$ and $2$.

In simulations of 2d quantum gravity using Markov Chain Monte Carlo methods the $N_m$ were
used as covariant observables for tracking thermalisation \cite{Surya:2011du}.
Importantly, it was observed that $N_m$ as a function of $m$ has a characteristic behaviour in the
phase in which flat spacetime is emergent, and differs drastically from the non-manifold
phase. Simulations of 2d flat spacetime showed that this characteristic curve could indeed be used
as a reliable indicator of flatness.

In this work we carry this idea forward. We begin by obtaining analytic expressions for the
expectation value of the interval abundances $\Nmd$ for a causal set that faithfully embeds into an Alexandrov
  interval in flat spacetime of arbitrary dimensions $d$.  Our main proposal is that the
characteristic curves for $\Nmd$ as a function of $m$ can be used to define a
local region in a manifoldlike causal set $C$. The existence of a local region in a causal set is
therefore a necessary condition for manifoldlikeness of $C$ and hence a new continuum dimension estimator. Specifically, since the characteristic
  curves for $\Nmd$ for fixed cardinality are sufficiently distinct for each $d$, it is possible to
  use them to find the continuum dimension of the local region in the causal set . This estimator
 therefore gives a null result for causal sets which are non-manifoldlike. Because the $\Nmd$ provide
an entire family of observables, it is tempting to conjecture that the requirement on interval
  abundances is not only a necessary but also a sufficient requirement for manifoldlikeness of a
causal set.

We test our proposals with simulations of causal sets that are approximated by spacetimes as well as
those that are not. We find that our necessary condition for manifoldlikeness works extremely well even for
relatively small causal sets. 

Indeed, not only do the simulated interval abundances reproduce on
average the characteristic curve, they follow it with reasonable precision even in a single realisation.

The latter is especially important in assessing manifoldlikeness in a single causal
set, as opposed to an ensemble of causal sets.  Our simulations verify that apart from being
able to determine the local regions of a manifold like causal set, our prescription is also a test
for manifoldlikeness itself and thence, manifold dimension.

Our construction demonstrates clearly the geometric richness of a locally finite poset which is
approximated by a spacetime. Our analysis  indicates the existence of  a local geometric ``rigidity'' present in
manifold like causal sets -- significant deviations from the $\Nmd$  result in Alexandrov intervals
that are explicitly {\it not} local. Thus, $\Nmd$ as function of $m$ provides us a local, covariant,
geometric measure for manifoldlikeness. 

In Section \ref{sec:preliminaries} we give a short introduction to the main concepts of causal set
theory and define quantities that we will need further on. The calculation of the interval
abundances in flat spacetime are in Section \ref{sec:flat}. We find that the ratio of these
abundances is scale invariant in the large $N$ limit, which provides a strong motivation for using
the $\Nmd$ as indicators of locality.  In Section \ref{sec:test} we present the main ideas in
  this work, namely how the $\Nmd$ can be used to define locality in a manifoldlike causal set. We
  then conjecture that the $\Nmd$ provide a rigidity criterion for manifoldlikeness for a causal set
  that faithfully embeds into an Alexandrov interval in flat spacetime.  In Section \ref{sec:sims}
we present results from extensive numerical simulations that support these ideas.
In particular, we examine the interval abundances of causal sets that we know a priori to be either
manifoldlike or not and find that our test works extremely well.  Our tests include flat spacetimes
of different dimensions, the $2$-d cut-trousers topology, as well as the flat geometries on $S^1
\times \re$ and $T^2 \times \re$, $4$-d FRW spacetimes, including deSitter spacetime, and some
examples of non-manifoldlike causal sets. In particular, we examine the claim from
\cite{Ahmed:2009qm} that causal sets grown with transitive percolation are manifoldlike. We find
that while macroscopic indicators may suggest manifoldlikeness, it fails our microscopic test.  We
end with a discussion on open questions and future directions in Section \ref{sec:conclusion}.

\section{\label{sec:preliminaries}Preliminaries}

Studies of Lorentzian geometry have long stressed the importance of the causal structure \cite{causalstructure}. 
For causal spacetimes, the causal structure provides a partial order on the
set of spacetime events. This partial order is a unique characteristic of a Lorentzian signature ($(-,+,+ \ldots
+)$) spacetime, a feature absent in  all other signature spacetimes.
It was shown by Malament, Hawking and others in \cite{malament:1399} that a bijection between two
past and future distinguishing spacetimes which preserves the causal structure is a conformal
isomorphism. Thus, knowing the causal relations between all points in a spacetime is enough to
define its geometry up to a conformal factor. The causal set 
hypothesis of a fundamental discreteness adds to this classical result by  
providing a discrete volume element to help recover  the conformal factor. Roughly, every discrete event comes
with approximately one unit of spacetime volume so that  the number of events
in a region corresponds to the volume of that region. In other words, in  causal set theory, an
appropriately discretised partially ordered set replaces continuum Lorentzian geometry, 
summarised in the slogan: Order + Number $\approx$ Spacetime.

Formally a causal set $C$ is defined to be a locally finite partially ordered set, namely a
countable set $C$ with an order relation $\p$ on its elements which is 

\begin{figure}
\includegraphics[width=0.45\textwidth]{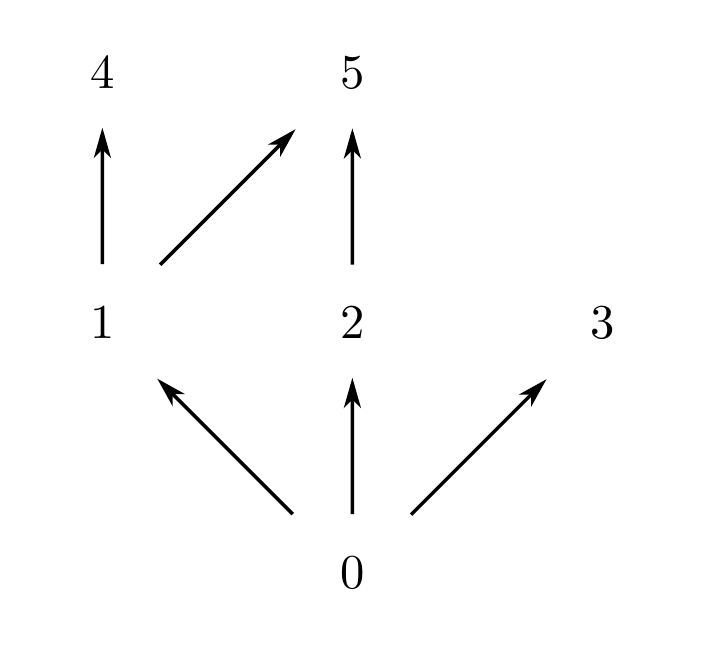}\hfill \includegraphics{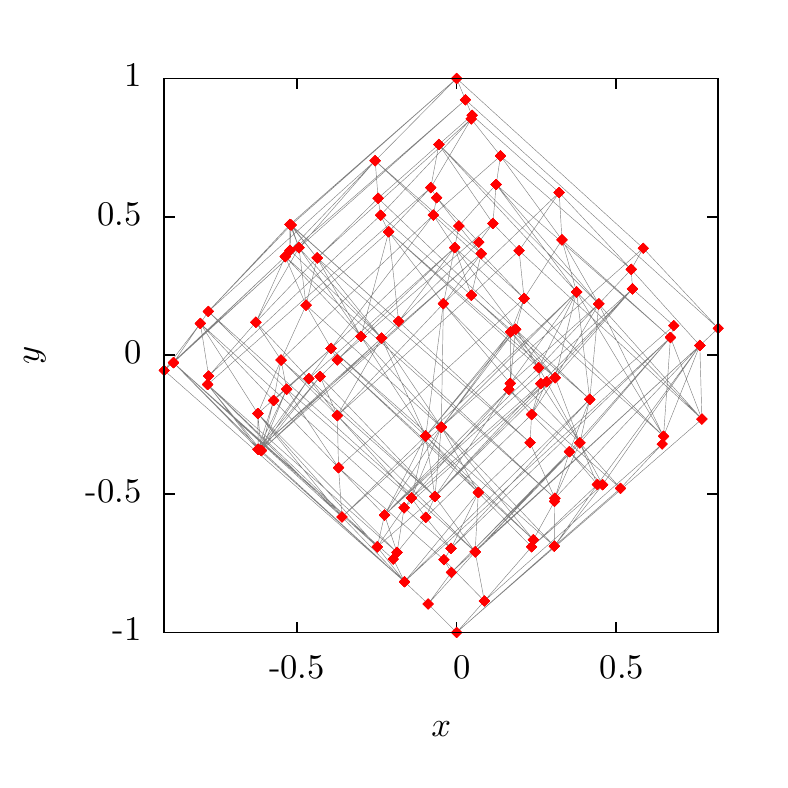}
\caption{\label{fig:causet}\small On the left is  the Hesse diagram of a simple causal set where
  only the links are shown by arrows. The figure on the right is a sprinkling of 100 elements  into
  an interval of flat spacetime, where all the causal relations are depicted by lines.} 
\end{figure}

\begin{itemize}
\item[(a)] {Reflexive:} for all $x
\in \C$, $x \p x$
\item[(b)] {Transitive:} for all $x,y,z \in \C$ and $x \p y$ and $y \p z$ then $x \p z$,
\item[(c)] {Acyclic:} for all $x,y \in \C$, $x \p y \p x \Rightarrow x =y$ 
\item[(d)] {Locally Finite:} for
all $x,y \in \C$ $| I(x,y)| \equiv |\{ z| x\p z \p y\}| < \infty $ \;. 
\end{itemize}
This last condition is equivalent to the assumption of a fundamental discreteness.  The first
  figure in Fig. \ref{fig:causet} shows the Hesse diagram of a small causal set
where the elements are numbered and the links are  denoted by 
arrows. 

Causal set quantum gravity is thus a quantum theory of causal sets with the continuum existing only
as an approximation to a fundamentally discrete substructure.  In particular, the ensemble of causal
sets that are approximated by a given spacetime $(M,g)$ is obtained via a Poisson process for a
given discreteness scale $\rho^{-1}$. The probability of assigning $m$-elements of a causal
set $C$ in a spacetime region of volume $V$ is given by
\begin{equation}\label{eq:poisson}
	P_V(m)= \frac{ (\rho V)^{m}}{m!} e^{-\rho V}.  
\end{equation}
The causal set is then recovered via the induced causal relations on the set of elements. This
Poisson ``sprinkling'' is a key feature of causal set discretisation of the continuum. The second 
figure in Fig.  \ref{fig:causet} shows a sprinkling of 100 elements into flat space. Conversely,
given a causal set, $C$, we say that it is approximated by a spacetime $(M,g)$ if there exists a
{\it faithful embedding} $\Phi: C \rightarrow (M,g)$ such that $\Phi(C) \subset (M,g)$ corresponds
to a high probability Poisson sprinkling into $(M,g)$ with the order relations in $C$ being the same
as those induced by the causal  relations in $(M,g)$ onto $\Phi(C)$.  An important conjecture in the theory is
that a given causal set cannot faithfully embed into two distinct spacetimes, namely, those  which differ on scales
larger than the discreteness scale.  In other words continuum structures below the discreteness
scale are irrelevant to the theory.  We refer the reader to the literature for more details on the
basics of the causal set hypothesis \cite{cst, valdivia, Dowker:2005tz,Surya:2011yh}.

As discussed in the introduction, the key focus of this work is to be able to define a local region
in a manifoldlike causal set. In the continuum, an Alexandrov interval $I[x,y] \equiv \{z| x\pprec
z \pprec y\} $, where $\pprec$ is the chronological relation. In a causal set, there is no a priori
distinction between causal and chronological relations and hence we define an {\sl order interval}
$I[x,y] \subset C$ as $I[x,y]=\{ z| x \prec z \prec y\} $.  A natural characterisation of $I[x,y]$
is its cardinality or discrete volume.  However, this
information does not suffice to distinguish an $I[x,y]$ which is local and one that is not. Since
the discrete geometry should include all relevant information about the continuum, we expect that
there must exist other observables in $I[x,y]$ which can be used to characterise locality.  In
this work, we find that the abundance of $m$-element order-intervals in $I[x,y]$ is indeed
such a family of observables. Namely for every $m$ we count the number or abundance of order
intervals of size $m$ in $I[x,y]$. For us, an $m=0$ order interval is a link, namely an order
  interval which contains only its end points, an $m=1$ order interval is one with a single element
  that lies between the end points, or an irreducible $3$ chain, and $m=2$ can be an irreducible $4$
  chain or an irreducible diamond poset, i.e.,  an order interval with two elements between the end
  points.

 For a causal set that faithfully embeds into flat spacetime, we find that the interval
   abundances follow a characteristic, monotonically decreasing curve as $m$ increases. It is
 this characteristic curve that we will use as a ``ruler'' to determine the locality of an
 order interval in a more general manifoldlike causal set. However, in order to do
   so, we would like to ensure that the scale of flatness in every region of the spacetime is much
   larger than the discreteness scale, so that the manifold approximation of the causal set is
   well-defined everywhere. By this we mean the following.  Consider a causal set that faithfully
   embeds into an ``approximately flat'' spacetime region in which Riemann normal coordinates are
   valid.  Such a region is characterised by a  dimensionless size which we will refer to as the {\sl
     scale of flatness} $\curv>>^{-1}>>1$ where $\curv=R\tau^2$ with $R$ denoting any component of
   the Riemann tensor at an event in the region and $\tau$ the proper time between any two events in
   the region. In flat spacetime $\curv =0$ and hence the size $\curv^{-1} \rightarrow \infty$ as
   expected. For a generic spacetime we  will refer to such regions as ``small'': for a given $R$, the size of the region
   $\tau$ must be small enough for $\curv <<1$.  Let $C$ be a causal set that faithfully embeds at
   density $\rho$ into an Alexandrov interval $I[x,y]$ of volume $V$ which lies in a region for which
   $\curv^{-1}>>1 $.  If $N \sim \rho V \sim 1 $ then the continuum approximation of $C$ breaks down.
   Thus, in order for the $I[x,y]$ to be adequately represented by the causal set, we require that
   $N>>1$,  i.e., the discreteness scale $N^{-1}$ must be small with respect to the scale of flatness
   $\curv^{-1}$.

\section{\label{sec:flat} The Abundances of Order Intervals in  Flat Spacetime} 

We now find closed form expressions for the abundance of order-intervals in a causal set $C$ that is
faithfully embedded into an Alexandrov interval $I[p,q]$ in flat
spacetime. To begin with we find the abundance of links $\langle N_0^d \rangle$. Lorentz
invariance then allows us to generalise this expression to that for $\Nmd$ in a straightforward 
way. Although we use series expansions to evaluate the integrals, the final
expressions take relatively simple closed forms.  

\begin{figure}
\centering
	\includegraphics{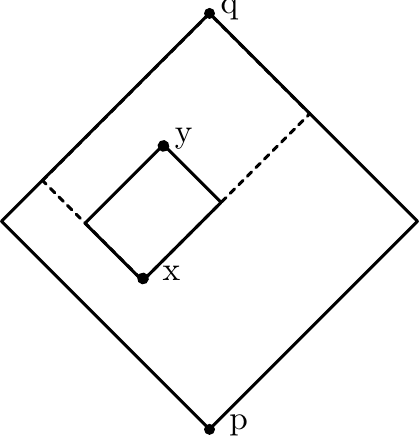}
	\caption{\label{fig:sketch}The Alexandrov neighbourhood $I[p,q]$ in the  integration
          Eqn.(\ref{eq:nolinks}): $x$ lies anywhere in $I[p,q]$ while $y$ lies in the intersection
          of $I[p,q]$ with the chronological future of $x$.}
\end{figure}

Consider the interval $I[p,q]$ in Fig. \ref{fig:sketch} with volume $V$ and proper time $\tau$ from
$p$ to $q$.  
The probability that there is a link
from an element $x$ to an element $y$ in this region is given by 
\begin{equation}
 	P_{xy} =  e^{-\rho V_{xy}},
\end{equation} 
where $V_{xy}$ is the spacetime volume of the Alexandrov interval  $I[x,y] \subset I[p,q]$
in the embedding spacetime. Moreover, {\it given} $x$, $y$ lies to its future, i.e., $y \in
I[x,q]$, while $x$ can lie anywhere in $I[p,q]$. Thus, the expectation value for the number of links
in $I[p,q]$ is given by
\begin{equation}\label{eq:nolinks}
  \langle N_{0}^{d}(\rho, V)\rangle= \rho^{2} \intl{\diamond}{} \md V_x \intl{\diamond_{x}}{} \md V_{y}
  e^{-\rho V_{xy}},
\end{equation}
where the symbol $\diamond$ denotes  $I[p,q]$ and the symbol $\diamond_{x}$ denotes $I[x,q]$.
Since this expression is Lorentz invariant, it can only depend on the proper time $\tau$ or volume
$V$ of $I[p,q]$. Thus we may choose $p$ at the origin $p=(0,\ldots, 0)$ and $q$ on the time
axis $q=(\sqrt a, 0, \ldots, 0)$, where  $\tau=\sqrt{2} a$.  

Lorentz covariance also implies that the integration over $\diamond_{x}$ depends only the proper
time $\tau(x,q)$ and the volume $V_{xq}$ of $I[x,q]$. We may thus again calculate this integral in
convenient coordinates and then recast it in terms of $\tau(x,q)$ and $V_{xq}$.  We take $x$ to lie
at the origin $x=(0,\ldots, 0)$ and $q$ to lie on the time axis $q=(\sqrt{2} a', 0, \ldots, 0)$,
where $\tau(x,q)=\sqrt 2 a' \leq \sqrt 2 a$.  Using lightcone coordinates
\begin{align}
	v&=\frac{1}{\sqrt{2}}(t+r) &u&=\frac{1}{\sqrt{2}}(t-r), 
\end{align}
the integration measure in flat space for  $y=(u_y,v_y, \vec \Omega_y)$ is then
\begin{equation}
	\int \md V_{y}= \frac{1}{2^{\frac{d}{2}-1}}\int \md \Omega_{y} \intl{0}{a'} \md v_y
        \intl{0}{v}\md u_y (v-u)^{d-2} \;.  
\end{equation}
The function we will integrate over does not involve the angular coordinates of $y$, since $\tau(x,y) =  \sqrt{2 u_yv_y} $.  We can then rewrite 
\begin{equation}
	\rho V_{xy} = \rho \frac{S_{d-2}\; 2^{-\frac{d}{2}+1}}{d (d-1)} (u_y v_y)^{\frac{d}{2}} =
        \rho \zeta_d  (u_y v_y)^{\frac{d}{2}} \;, 
\end{equation}
where $S_{d-2}$ is the volume of the $d-2$ sphere $ S_{d-2}= \int \md \Omega_y = \frac{2
  \pi^{\frac{d-1}{2}}}{\G{\frac{d-1}{2}}}, $ and we define the dimension dependent constant $
\zeta_d \equiv \frac{S_{d-2}\; 2^{-\frac{d}{2}+1}}{d (d-1)}. $ Thus the angular integration over
$\diamond_{x} $ factors out so that, in these coordinates, the integral $I'_{\diamond_x} \equiv
\intl{\diamond_{x}}{} \md V_{y} e^{-\rho V_{xy}}$ reduces to
\begin{equation} 
I'_{\diamond_x} = d(d-1) \zeta_d \intl{0}{a'} \md v \intl{0}{v}\md u (v-u)^{d-2} e^{-\rho \zeta_d (u
  v)^{d/2}}, 
\end{equation} 
where we have suppressed the subscripts in $(u_y,v_y)$. Expanding $(v-u)^{d-2}$ in terms of binomial coefficients and $e^{-\rho
  \zeta_d (u v)^{d/2}}$ as a power series simplifies the integration considerably
\begin{align} 
I'_{\diamond_x} = &  \, d(d-1) \zeta_d \suml{n=0}{\infty} \frac{\left(-\rho\zeta_d \right)^{n}
        }{n!} \suml{k=0}{d-2}\binom{d-2}{k} (-1)^{k} \intl{0}{a'} \md v \intl{0}{v}\md u \;
        v^{d(\frac{n}{2} +1)-2-k} u^{\frac{d n}{2}+k} \\
	=& \, d(d-1) \zeta_d\suml{n=0}{\infty} \frac{\left(-\rho\zeta_d \right)^{n} }{n!}
        \frac{a'^{d(n+1)}}{d(n+1)} \suml{k=0}{d-2}\binom{d-2}{k} \frac{(-1)^{k}}{\frac{d
            n}{2}+k+1}. 
\end{align} 
Rewriting 
\begin{equation}\label{eq:binsum}
	\suml{k=0}{d-2}\binom{d-2}{k} \frac{(-1)^{k}}{\frac{d n}{2}+k+1}= \frac{\G{d-1} \G{\frac{dn}{2}+1}}{\G{\frac{d}{2}(n+2)}}
\end{equation}
we find that 
\begin{equation} 
I'_{\diamond_x}	= \, d(d-1) \zeta_d 
\G{d-1}  \suml{n=0}{\infty} \frac{\left(-\rho\zeta_d \right)^{n} }{n!} \frac{a'^{d(n+1)}}{d(n+1)} \frac{\G{\frac{dn}{2}+1}}{\G{\frac{d}{2}(n+2)}}
\end{equation} 
We can now convert the above expression into a Lorentz covariant form, by substituting $a'$ for the
proper time $\tau(x,q)=\sqrt{2} a'$ of $I[x,q]$. In the original coordinates adapted for
$I[p,q]$ this is  $\tau(x,q)^{2}= 2 (a-v_x)(a-u_x)$    
Thus, to complete the calculation of  $\Nzerod$ we must evaluate the integral 
\begin{equation} 
I_\diamond=\intl{\diamond}{} \!\!\md V_x 
((a-u_x)(a-v_x)) ^{\!\!\frac{d(n+1)}{2}} \! \! \!= \zeta_d d(d-1) \intl{0}{a} \md v_x \intl{0}{v_x}
\md u_x (v_x-u_x)^{d-2}  ((a-u_x) (a-v_x) )^{\frac{d(n+1)}{2}} .   
\end{equation} 
To shorten the notation we defined this integral without the sum over $n$ which we will have
  to restore in the final expression for $\langle N_{0}^{d} \rangle $.
Substituting $u=a-u_x $ and $v= a -v_x$ and again using the binomial expansion  
\begin{eqnarray} 
 I_\diamond & = & \zeta_d d(d-1)  \suml{k=0}{d-2}\binom{d-2}{k} (-1)^{k} \intl{0}{a} \md v \intl{0}{v} \md u \; u^{\frac{d}{2}(n+1)+k}  v^{\frac{d}{2}(n+3)-2-k} \\
  &=& \zeta_d d(d-1)\frac{a^{d(n+2)}}{d(n+2)}  \suml{k=0}{d-2}\binom{d-2}{k} (-1)^{k}
  \frac{1}{\frac{d}{2}(n+1)+k+1}.  
\end{eqnarray} 
Using the identity \eqref{eq:binsum}, and the fact that $V=\zeta_d a^{d} $, we find the Lorentz
covariant expression for the average number of links in an interval $I[p,q]$ of volume $V$ to be
\begin{equation}
  \Nzerod(\rho,V)=	\G{d}^{2}  \suml{n=0}{\infty} \frac{\left( -\rho V \right )^{n+2}}{(n+2)!} \frac{\G{\frac{dn}{2}+1}}{\G{\frac{d}{2}(n+2)}} \frac{\G{\frac{d}{2}(n+1)+1}}{\G{\frac{d}{2}(n+3)}} \;.
\end{equation}

This expression can now be used to find  the  $m$ element interval abundances in flat space
for  general $m$ by observing that \eqref{eq:poisson} can be rewritten as 
\begin{equation}\label{eq:genm}
	P(m,V,\rho)= \frac{\left( -\rho\right)^{m}}{m!} \pd{^{m}}{\rho^{m}} e^{-\rho V} \;.
\end{equation}
Thus, the average  number of $m$-element  intervals in a volume $V$ is simply given as 
\begin{eqnarray} 
\label{eq:deff}
	\Nmd(\rho, V) &=&  \rho^{2} \intl{\diamond}{} \md \vy \intl{\diamond_{y}}{} \md V_{x}
        \frac{ (\rho V)^{m}}{m!} e^{-\rho V} \nonumber \\ &= & 
	\frac{\left( -\rho\right)^{m+2}}{m!} \pd{^{m}}{\rho^{m}} \intl{\diamond}{} \md V_{x}
        \intl{\diamond_{y}}{} \md \vy\; e^{-\rho V}  \nonumber \\
        &=&\frac{ (-\rho)^{m+2}}{m!} \pd{^{m}}{\rho^{m}}\rho^{-2} \Nzerod(\rho,V)
\end{eqnarray} 
Which evaluates to:
\begin{eqnarray} 
\Nmd(\rho,V) & = & \frac{\G{d}^{2} }{ m!} \left( -\rho V \right )^{m+2} \suml{n=0}{\infty}
\frac{\left( -\rho V \right )^{n}}{n!}\frac{1}{(n+m+1)(n+m+2)} \times \nonumber \\ && 
\frac{\G{\frac{d}{2}(n+m)+1}}{\G{\frac{d}{2}(n+m+2)}} \frac{\G{\frac{d}{2}(n+m+1)+1}}{\G{\frac{d}{2}(n+m+3)}} \;.
\end{eqnarray} 
This can then be recast as  a closed form expression in terms of  generalised hypergeometric
functions:
\begin{align}
	\Nmd(\rho,V)=& \frac{(\rho V)^{m+2}}{(m+2)!} \frac{\G{d}^{2}}{\Poch{\frac{d}{2} (m + 1) + 1}{ d - 1}} \frac{1}{\Poch{\frac{d}{2} m + 1}{d - 1}} \nonumber \\
	& \mFm{d}{1+m,\frac{2 }
	{d}+m,\frac{4}{d}+m, \dots ,\frac{2 (d-1)}{d}+m)}
	{3+m,\frac{2 }{d}+m+2,\frac{4}{d}+m+2, \dots ,\frac{2 (d-1)}{d}+m+2}
	{- \rho V} \;,
\label{eq:flatclosedform} 
\end{align}
where $_{p}F_{q}(\{a_1, \ldots, a_p \}, \{b_1, \ldots, b_q\}|-z)$ is a generalised hypergeometric
function and $(a)_n$ is the Pochhammer symbol.
This expression is convergent because, as is well known, generalised hypergeometric functions
converge for all $z$ values if $p \leq q$.  The details of obtaining this
form for the $\Nmd$ are given in appendix \ref{app:closedform}.

In Fig. \ref{fig:profiles} we plot the function $\Nmd$ for different values of $d$. $\Nmd$
rapidly and monotonically decreases as $m$ increases thus providing  a clear characteristic
signature for the flat spacetime case, which we will use to define locality and thence a
  continuum dimension estimator.
\begin{center} 
\begin{figure}
\center
	\includegraphics{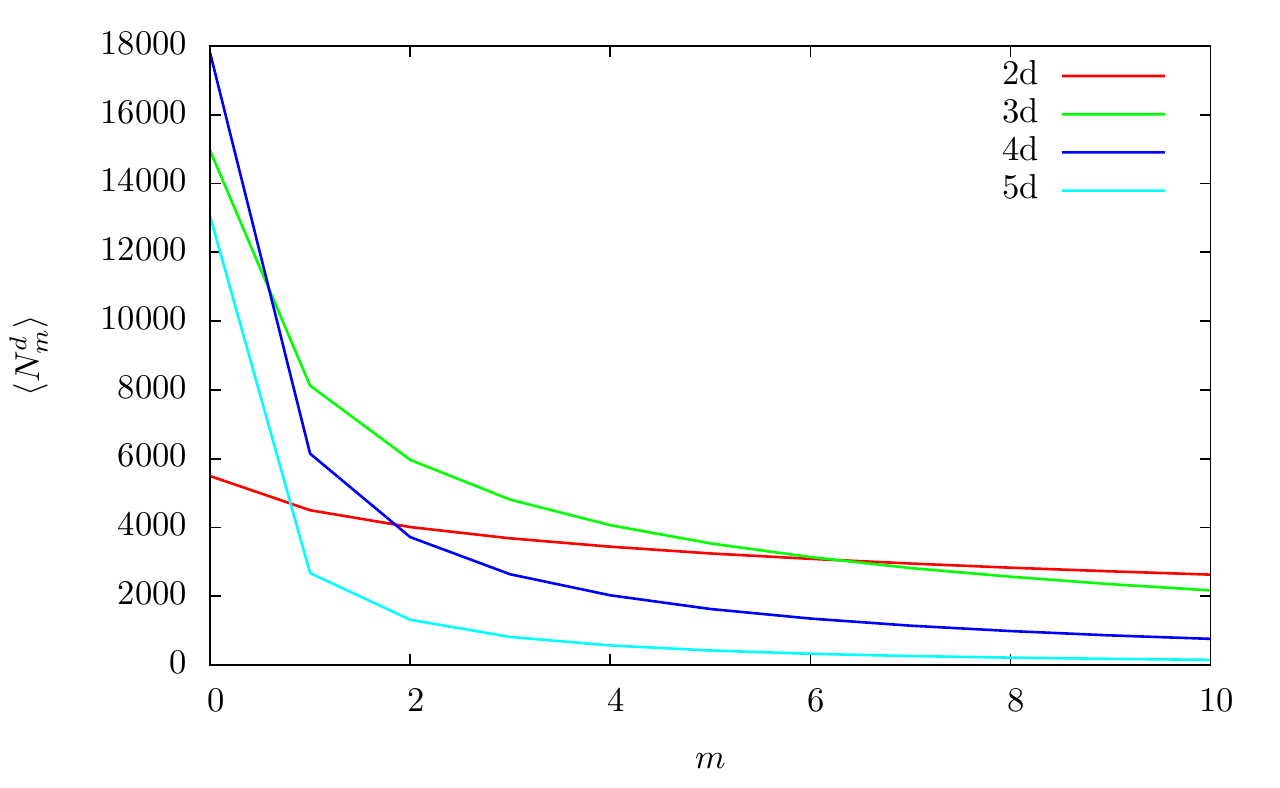}
\caption{\label{fig:profiles} The function $\Nmd$ v/s $m$  for $N=1000$ and  $d=1,\ldots, 5$.}
\end{figure}
\end{center} 

\subsection{\label{continuum}  The Asymptotic Limit} 

We now examine the behaviour of these expressions in the large $N=\rho V$ limit. Although the
continuum limit is not physically relevant per se to causal set theory, it is nevertheless an
interesting limit of the theory.  Clearly, $\Nmd$ will diverge with $N$, but it is not clear apriori
what the behaviour will be after normalisation, say with respect to the abundance of links,
$\Nzerod$. If there were a leading order $N$-dependence, then either this would diverge, or go to
zero in the limit, thus drastically changing the nature of the $\Nmd$ characteristic curve.

What we find is that the ratio is in fact  independent of $N$ to leading order and yields a
surprisingly simple expression  in the asymptotic limit 
\begin{equation} \label{eq:limitdeftwo}
\Sc_m^d \equiv \lim_{\rho \rightarrow \infty} \frac{\Nmd(\rho, V)}{\Nzerod(\rho, V)}=\frac{
  \G{\frac{2}{d}+m} }{\G{\frac{2}{d}} \G{m+1}}\; . 
\end{equation}
This scale invariance seems to echo that of Minkowski spacetime suggesting that the $\Nmd$ captures
an essential and perhaps even defining ingredient of flat spacetime geometry. We discuss this in some
detail in the following section.

Here we give a quick sketch of how this limit is obtained, leaving details to appendix \ref{app:continuum}. The
$N=\rho V$ dependence in Eqn. (\ref{eq:flatclosedform}) comes from the overall $N^{m+2}$ factor as
well as the hypergeometric function $_dF_d$ which when appropriately rearranged is of the form
\begin{align}\label{eq:shapetwo}
\mFm{d}{a_1, \dots, a_d }{a_1+2,\dots , a_d+2}{-N} \;, \quad a_i=\frac{2i}{d} + m, \, \,i=1, \ldots d-1,
\quad  a_d=1+m.
\end{align} 
Thus, to investigate the $N\to\infty$ limit of the $\Nmd$ we need a large $N$ expansion of
this function.  We make repeated use of the following identity \cite{wolfram1}
\begin{align}
  &\, \pFq{p}{q}{a_1,\ldots ,a_p}{a_1+m_1,\ldots ,a_n+m_n,b_{n+1},\ldots ,b_q}{z}= \prod _{j=1}^n \frac{\left(a_j\right)_{m_j} }{\left(m_j-1\right)!} \sum _{k=1}^n \sum _{j_1=0}^{m_1-1} \ldots  \sum _{j_n=0}^{m_n-1} \frac{1}{a_k+j_k} \nonumber \\
  &\prod _{l=1}^n \frac{\left(1-m_l\right)_{j_l}}{j_l!} \prod _{\substack{i=1\\ i\neq k}}^n
  \frac{1}{a_i+j_i-a_k-j_k} \, \pFq{p-n+1}{q-n+1}{a_k+j_k,a_{n+1},\ldots
    ,a_p}{a_k+j_k+1,b_{n+1},\ldots ,b_q}{z},  \label{eq:pFq} \\
  &\ \ \ \ \ \ \ \ \ \ \ \ \ \ \ \ m_n\in \mathbb{Z}\land m_n>0\land n\leq q \land a_{i} + j_i \neq a_k+ j_k,
  \forall j_i=0,.. \ldots m_i, 1 \leq i,j \leq n\nonumber.
\end{align}
This can be used to reduce the $_dF_{d}$ of the form in Eqn. (\ref{eq:shapetwo}) to (i) a sum over
$_1F_{1}$ in odd dimensions, (ii) a sum over $_3F_{3}$ in even dimensions $d>2$. In $d=2$ $_dF_{d}$
is simply $_2F_{2}$ which can be examined directly.  We demonstrate these results explicitly in
appendix \ref{app:continuum}.

Specifically, in odd dimensions there is a sum over  $\mFm{1}{a_k+j_k}{a_k+j_k+1}{-N}$, which using  
\begin{align}\label{eq:1F1}
\, \mFm{1}{a}{a+1}{-z}&=a(z)^{-a} (\Gamma (a)-\Gamma (a,z)) \\
\G{a,z}&\propto e^{-z} z^{a-1} \left(\frac{(2-a) (1-a)}{z^2}-\frac{1-a}{z}+\ldots +1\right)\text{/;}(\left| z\right| \to \infty ) \;. \label{eq:Gaz}
\end{align}
gives a leading order dependence of $N^{-a_k-j_k}$, for the smallest values of $a_k+j_k$ which is
$k=1$, $j_1=0$, which makes it $N^{-\frac{2}{d}-m}$.  In even dimensions for $d>2$, the dependence on $N$
appears in a sum over
\begin{equation} 
\mFm{3}{\frac{2}{d}k+m+j_k, \frac{2}{d} l +m+1, m+1}{\frac{2}{d}k+m+j_k+1, \frac{2}{d}l+m+2,
  m+3}{-N}.
\end{equation}      
We obtain an asymptotic expansion of this using Mathematica and find that the leading order
contribution is again $N^{-a_k-j_k}$ and hence comes from the $k=1, j_1=0$ term which makes it $\sim
N^{-\frac{2}{d}-m}$.  In $d=2$, the hypergeometric function in Eqn. (\ref{eq:flatclosedform}) is
simply $\mFm{2}{m+1,m+1}{m+3,m+3}{-N} $ whose leading order contribution is of the form $N^{-1-m}
\log N$.  Combining these we find that $\Nmd \sim N^{2-2/d}$ to leading order for $d>2$ and $\Nmd
\sim N \log N$ for $d=2$.
 
What our detailed calculations show, moreover, is that in {\it all} dimensions  the
coefficient of the leading order term takes the simple form:
\begin{equation} 
\frac{1}{m!} \G{\frac{2}{d}+m} \frac{\G{d} }{ \left(\frac{d}{2} -1\right) \Poch{\frac{d}{2}+1}{d-2}} 
\end{equation} 
which implies Eqn. (\ref{eq:limitdeftwo}).  The subleading contributions however vary from
dimension to dimension as 
\begin{equation}
 \Nmd(N) =\frac{N^{2-\frac{2}{d}}}{m!} \G{\frac{2}{d}+m} \frac{\G{d} }{ \left(\frac{d}{2} -1\right) \Poch{\frac{d}{2}+1}{d-2}}+ \begin{cases}
 \mathcal{O}(N ) & \text{ for } d=3\\
 \mathcal{O}(N \log{N}) & \text{ for } d=4\\
 \mathcal{O}(N^{2-\frac{4}{d}}) &\text{ for } d>4 
 \end{cases}
\end{equation}
for all $d>2$ and 
\begin{equation} 
\Nmdv{m}{2}(N)= N \log{N} + \mathcal{O}(N) \;,
\end{equation} 
for $d=2$.  We refer the reader to appendix \ref{app:continuum} for the details of the calculation. 

\begin{figure}
\subfloat[Convergence in 4d]{ \includegraphics[width=0.5\textwidth]{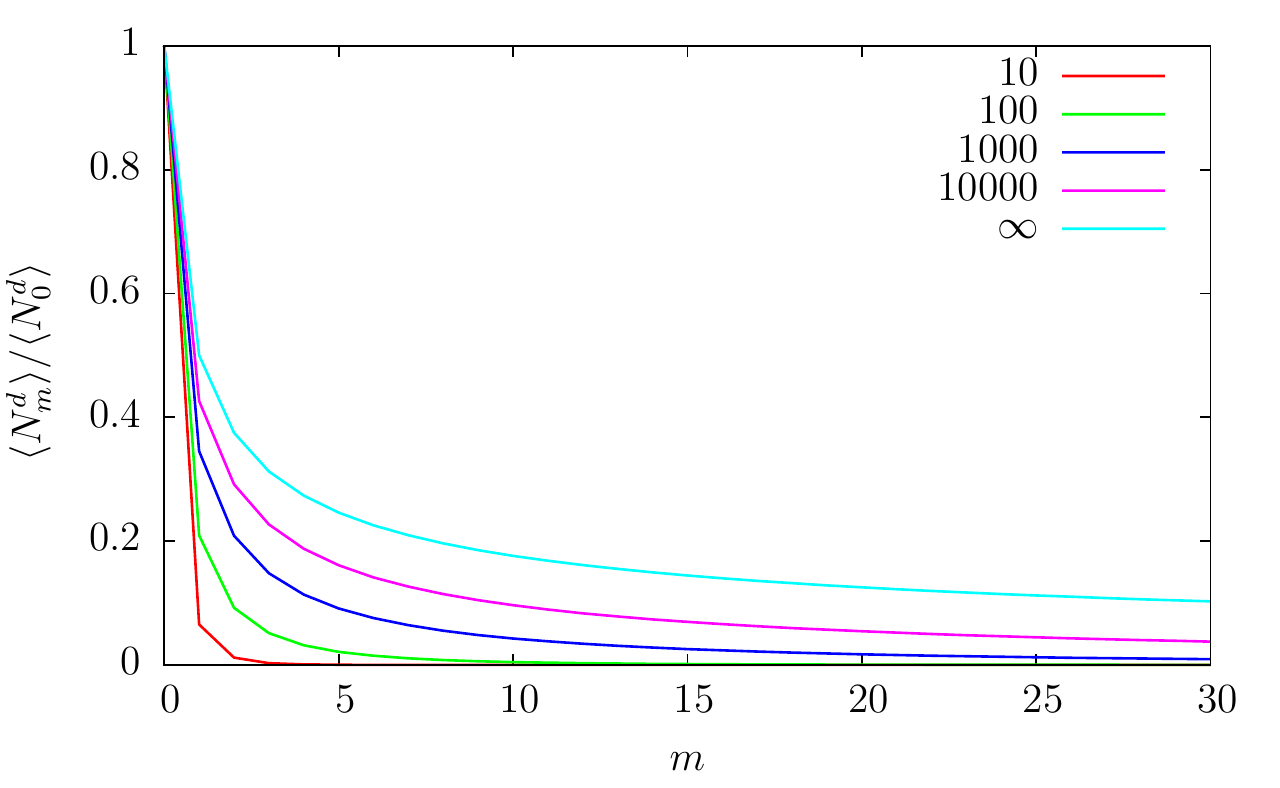}  }
\subfloat[Varying Dimension]{ \includegraphics[width=0.5\textwidth]{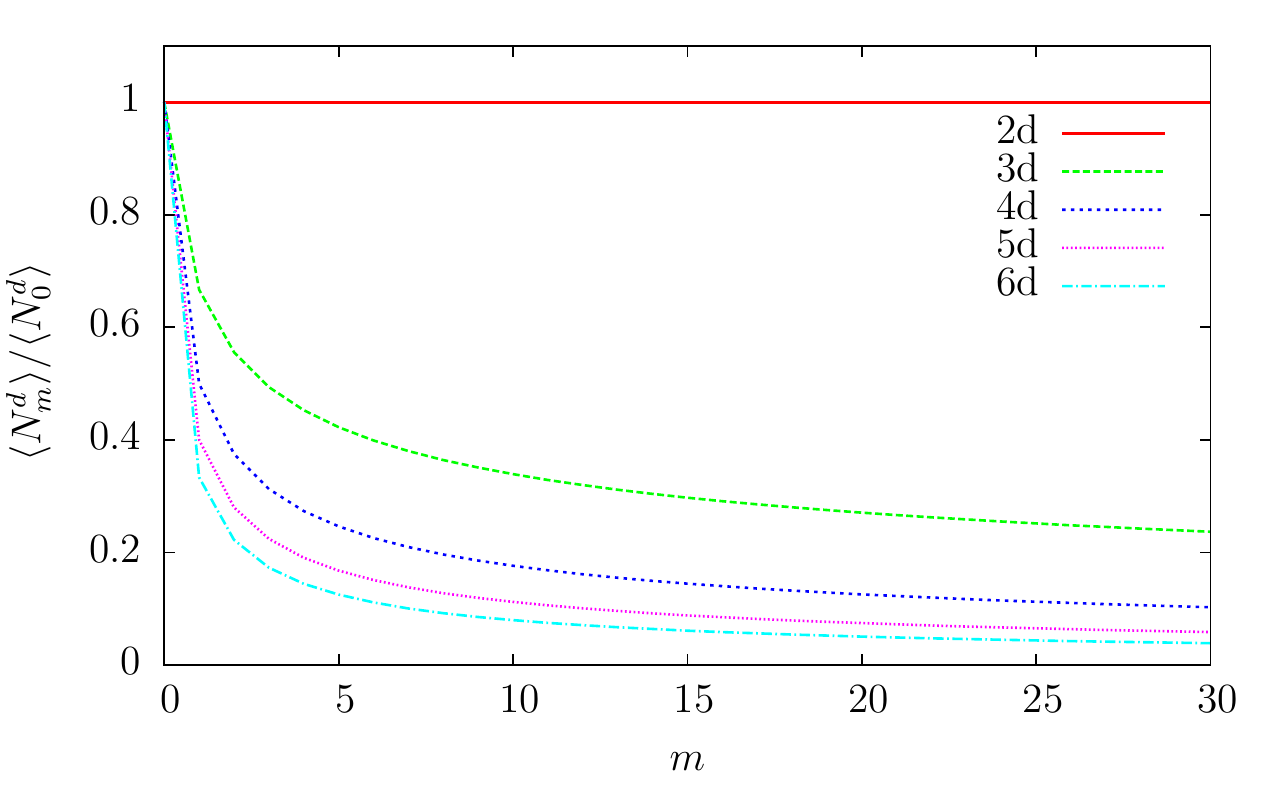}  }
\caption{\label{fig:limitplots}These plots illustrate some properties of the $N \rightarrow \infty$
  limit. The left hand plot shows the convergence of $\frac{\Nmd}{\Nzerod}$ to   $S_m^d$ in $4$d for
  $N=10,100,1000,10000$. The right hand plot shows  how $S_m^d$ changes  with dimension. }
\end{figure}

In particular we note that all contributions are slower than $N^2$ and that the convergence towards
the limit happens polynomially, and hence is quite slow.  In Fig. \ref{fig:limitplots} we plot
$\Nmd(N)$ for $d=4$ for a range of $N$-values, as well as the asymptotic limit. The
slow convergence makes it clear that it will not be possible to test this limit computationally.
In Fig. \ref{fig:limitplots} we plot the asymptotic limits for various $d$. Notably, for $d=2$
$S_m^2=1$ and therefore  independent of $m$. 

\section{\label{sec:test} Defining Local Regions in a  Causal Set}

The main goal of this work is to show that $\Nmd$ can be used as a definition of locality in a
causal set which faithfully embeds into a continuum spacetime. Conversely it can be used as a test
for manifoldlikeness as well as a continuum dimension estimator. 

Consider an $N$-element causal set $C$ which faithfully embeds into an Alexandrov interval $I[p,q]$
in $d$-dimensional Minkowski spacetime at a given density $\rho$. As we have just shown above, if
one considers the ensemble of causal sets obtained via a Poisson sprinkling into $I[p,q]$, then the
average $\Nmd$ has a characteristic behaviour with $m$. For large enough $\rho$, the
interval abundances $\NmCC$ for a single ``typical'' realisation will with high probability
``track'' $\Nmd$, i.e., $\NmCC \sim \Nmd(N \pm \sqrt N) $ for all $m$. This is what we would expect
from a Poisson distribution.  As we will show in the following section, this expectation is
confirmed by simulations. Simulations moreover show that the distribution of
  the $N_m^d$ for any given $m$ for  an ensemble of causal sets obtained via a Poisson sprinkling
  into $I[p,q]$  is nearly  Gaussian with a standard deviation of $\sim \sqrt{N}$.

Importantly, the closeness of a typical $\NmCC$ to $\Nmd$ can be used
as a characterisation of locality.  Namely, if $C$ is such that $\NmCC \sim \Nmd(N \pm \sqrt
N) $ for all $m$, and for a fixed $d$ we will refer to it as a ``local'' causal set. For a causal set $\tC$ which
faithfully embeds into an Alexandrov interval $I[p,q]$ in an arbitrary curved $d$-dimensional
spacetime, one expects that because of the deviation from flatness, $\NmtC$ will differ
significantly from the $\Nmd$. Again, this is borne out by simulations. Thus, $\tC$ is ``non-local'' in
this sense. However, as long as the scale of flatness  everywhere in $I[p,q]$ is much larger than
the discreteness scale as discussed in Section\ref{sec:preliminaries}\footnote{We will henceforth
  always assume that this condition is met.}, $\tC$ will contain $N$-element sub-causal sets $C$
which lie in an approximately flat Alexandrov interval $I[p',q'] \subset I[p,q]$. If $N$ is large
enough, then $C$ will be local in the above sense. Thus, the $\Nmd$ provide a strong
characterisation of local regions in a causal set. Again, this is borne out by simulations on a
class of curved spacetimes as well as those with non-trivial topology.

Thus, the function $\Nmd$ suggests a criterion for ``rigidity'' of $C$ in the sense used by
mathematicians. Namely, if $\NmCC  \sim \Nmd(N \pm \sqrt N) $ for all $m$, then it
suggests that $C$ must faithfully embed into Minkowski spacetime of dimension $d$ at large enough
embedding density. We now formalise these ideas as best as we can, leaving a more detailed study to
future work.

\begin{definition} 
  We will say that an $N$-element causal set $C$ is {\bf strongly $d$-rigid} if $\exists$ a
  $d$ for which $\NmCC \sim \Nmd(N \pm \sqrt N)$.  If $C$ possesses an $N'$ element sub causal
  set $C'$ which is strongly $d$-rigid, then $C$ is said to be {\bf weakly $d$-rigid}
  with respect to $C'$.
\end{definition} 

Clearly, strong $d$-rigidity is a necessary condition for an $N$-element causal set $C$ to
faithfully embed into an Alexandrov interval of flat $d$-dimensional spacetime of volume $V$ as long
as $N>>1$. On the other hand, weak $d$-rigidity is a rather weak necessary condition for $C$
to faithfully embed into a $d$-dimensional curved spacetime, since the only requirement
is that there exist a local or strongly $d$-rigid sub-causal set $C'$ in $C$. Indeed, in this
case, one should expect a whole family of strongly $d$-rigid sub causal sets $\{ C_i'\}$ in
$C$ for fixed $d$. However, a straightforward analysis of this case is far from clear at the
  moment and we leave this for future investigations.

We summarise the above in the following Claim: 

\begin{claim} \label{claimone} Let $C$ be an $N$-element causal set that faithfully embeds into an
  Alexandrov interval $I[p,q]$ in a $d$-dimensional spacetime such that the discreteness
    scale is much smaller than the scale of flatness everywhere. Then there exists a sub causal set
       $C' \subset C$ of cardinality $N'>>1$ such that $C'$ is strongly $d$-rigid. Moreover,
     if $I[p,q]$ is an Alexandrov interval in $d$-dimensional Minkowski spacetime then for large
  enough $N$, $C$ is itself strongly $d$-rigid.
 \end{claim}

\vskip0.2cm 

While the above arguments require that $N$ and $N'$ be arbitrarily large in order to suppress
 fluctuations, the simulations that we will present in the next section show that the necessary
 condition works extremely well even for $N'$ values as low as $100$ for a {\it single}
 ``typical'' realisation of $C'$.  This is true both for the flat spacetime case as well as for
 regions where the  scale of flatness is large .  Of course, for a generic curved
 spacetime, one does need to go to higher densities, but here too, there is strong evidence that the
 numbers can be relatively small.

 Could this condition also be sufficient for manifoldlikeness? As discussed above, in the general
 case, it clearly is insufficient since one needs requirements on an appropriately chosen family of
 strongly $d$-rigid sub-causal sets in $C$. On the other hand, it is a plausible sufficiency
 condition for a causal set to be faithfully embeddable into an interval in $d$-dimensional
 Minkowski spacetime.  There are several hints that support this. We first note that the interval
 abundance profile for generic causal sets or Kleitman Rothschild posets \cite{KR} which dominate
 the class of posets for large $N$ differs vastly from $\Nmd$. This difference in profile is easy to
 understand: these posets have a large number of links but almost no two or three element
 intervals. Thus, even at relatively small $m$, the interval abundances differ drastically from
 $\Nmd$.  Similar arguments can be made for the multiple layered class of causal sets studied in
 \cite{Dhar1,Dhar2} which are sub-dominant but are also largely devoid of small intervals with
 $m>1$.  In Fig.  \ref{fig:NonManifold} we show the interval abundances for a chain and a KR
 poset. Another example of a non manifoldlike causal set are the $2D$ orders corresponding to the
 crystalline phase of \cite{Surya:2011du}.  These are again layered, much like the KR posets, but
 here too, there is a large deviation from the flat spacetime $\Nmd$ .  Of course such examples
 cannot suffice since the space of causal sets is littered with those that have no simple
 characterisation.  Hence we cannot at the moment prove that strong $d$-rigidity is violated for
 {\it all} causal sets which do not faithfully embed into an interval in flat spacetime.

On the other hand, as shown in Section \ref{continuum} the ratio $\langle N_m^d \rangle/\langle
N_0^d\rangle$ is scale invariant in the limit $N \rightarrow \infty$.  In particular, this  mimics the
scale invariance of flat spacetime. Prompted by  discussions with Sorkin we   conjecture: 
\begin{conjecture} 
  If the interval abundances $\NmCC$ for an $N$-element causal set $C$ are such that that   $\NmCC
  \sim \Nmd(N\pm \sqrt N)$ for some $d$ in the large $N $ limit, then $C$
  faithfully embeds into an Alexandrov interval in $d$-dimensional Minkowski spacetime.
\end{conjecture} 
In other words, we suggest that $\Nmd $ provides a rigidity condition for a causal set {to be}
approximated by an Alexandrov interval in Minkowski spacetime.  A continuum version of this would
require $C$ to moreover be scale invariant or homogeneous, and it would be interesting to explore
whether there are examples of homogeneous orders like the Box spaces \cite{brightwellbolobas} which
could provide counter-examples to the conjecture \footnote{We thank Rafael Sorkin for
    discussions on the continuum limit.}.

\section{\label{sec:sims}Simulations} 

We now show evidence for the above results and conjectures using simulations of relatively small
causal sets.  We consider causal sets that are sprinkled into flat and curved spacetime as well as
non manifoldlike causal sets, using the Cactus-Code causal set framework
\cite{dr-cactus,cactus2}. In particular, we perform our test on causal sets discretisations of flat
spacetime for $d=2,\ldots, 4$ as well as on the $2$-d cut-trousers and the flat geometries on $S^1
\times \re$ and $T^2 \times \re$ .  As examples of curved spacetime we consider FRW spacetimes for
$d=4$ including deSitter spacetime,  both for small and large scales of flatness,  and find
significant deviations from the flat spacetime curves in the case of a small scale of flatness. 
All these examples provide ample support for Claim \ref{claimone}

\ref{claimone} 
even for relatively small $N$.  Next, we consider simulations of causal sets generated by
transitive percolation for the specific cases studied in \cite{Ahmed:2009qm} and show that they do
not pass our test for manifoldlikeness.  Finally, as support for our conjecture, we examine
distinctly non-manifoldlike causal sets, a chain and the class of Kleitman-Rothschild causal sets
and show that, as expected, they fail our test of manifoldlikeness.

Once the causal set $C$ is simulated, the interval abundances can be obtained within an
appropriately chosen order interval $I[p,q] \subseteq C$.  We employ two different procedures 
for this purpose.  The first procedure is a test of locality of an entire
  causal set. Here we consider sprinklings into a large interval in flat
spacetime and ``cap'' $C$ to the past and the future by adding a pair of extra elements $p,q$ so
that $I[p,q]=C$. This allows us to measure the interval abundance for the entire causal set. It is
especially useful when comparing the results from simulations into flat spacetime with the analytic
plots. Thus, we do not look for local regions in a given causal set, but test for the locality
  of an entire causal set, or in the language of the previous section whether it is strongly
  $d$-rigid for some $d$.

  The second procedure is for finding local regions in a causal
    set $C$ which may not itself be local.Here we pick out an element ${p} \in
  C$ and then examine the set of order intervals to which $p$ belongs. By comparing with the
  $\Nmd$ curves, one can then identify which of these order intervals might serve as a local
  neighbourhood of $p$.  This method has two hurdles we must overcome.

Firstly, we can not control the location of $p$  in the embedding
spacetime. This makes it hard to find intervals that sample a specific feature of a spacetime, say a
singularity in the cut-trousers topology of \ref{sub:cut} or a point close to the origin of the FRW
spacetime. A little control can be exerted using the fact that the current Cactus code uses a
natural labelling of the causal set, namely if $p\prec q $ then the labels satisfy $l(p) < l(q)$.
Thus, picking a point with a low/ high labelling allows us to choose the lower/upper area of the
region we sprinkled in.

Secondly, the number of intervals that contain an element can be very large even for
moderately sized causal sets. This can be ameliorated by only examining intervals
within a certain size range.

Thus our second procedure will be  to pick an
appropriate element in the causal set and then examine all intervals, within a certain size
range, that contain this element.

\subsection{Flat spacetime simulations}
\begin{figure}
{\centering 

\subfloat[2d -100 Points]{\includegraphics{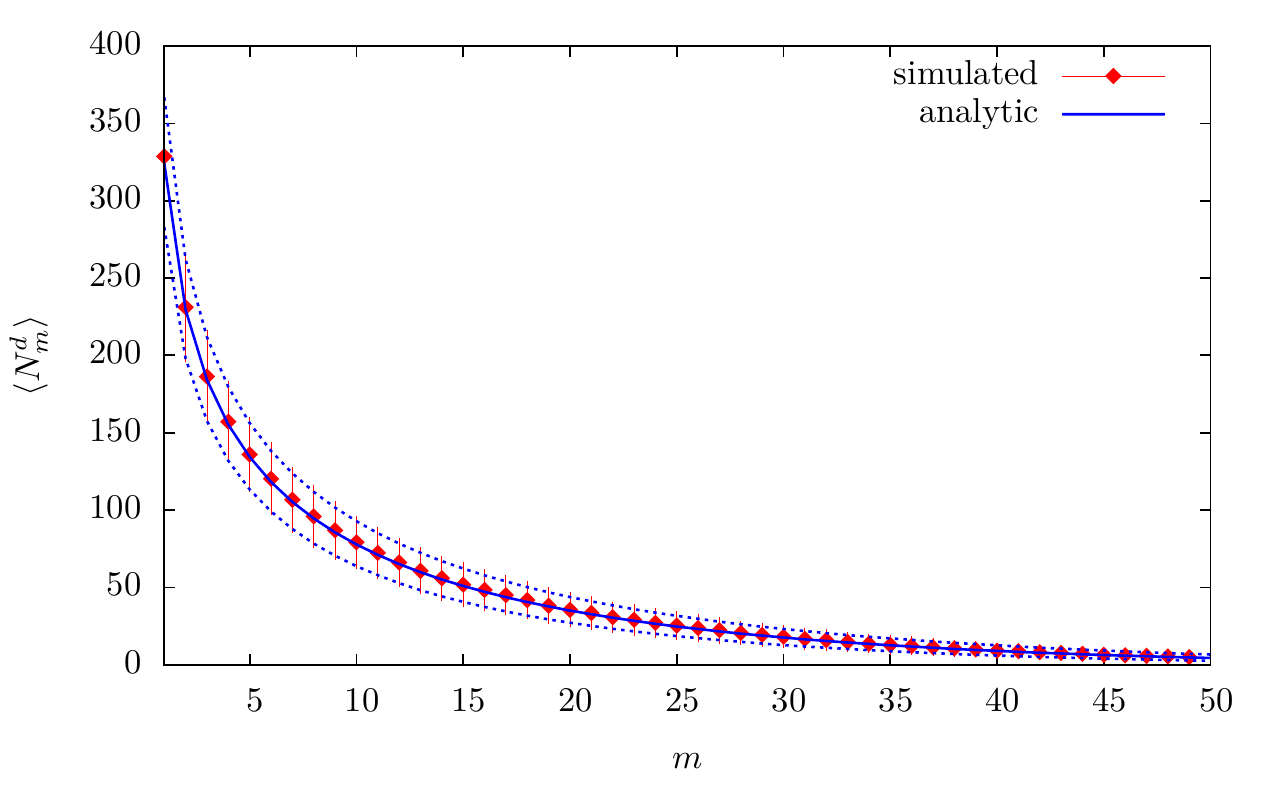} }

}

\subfloat[3d - 1000 Points]{\includegraphics[width=0.5\textwidth]{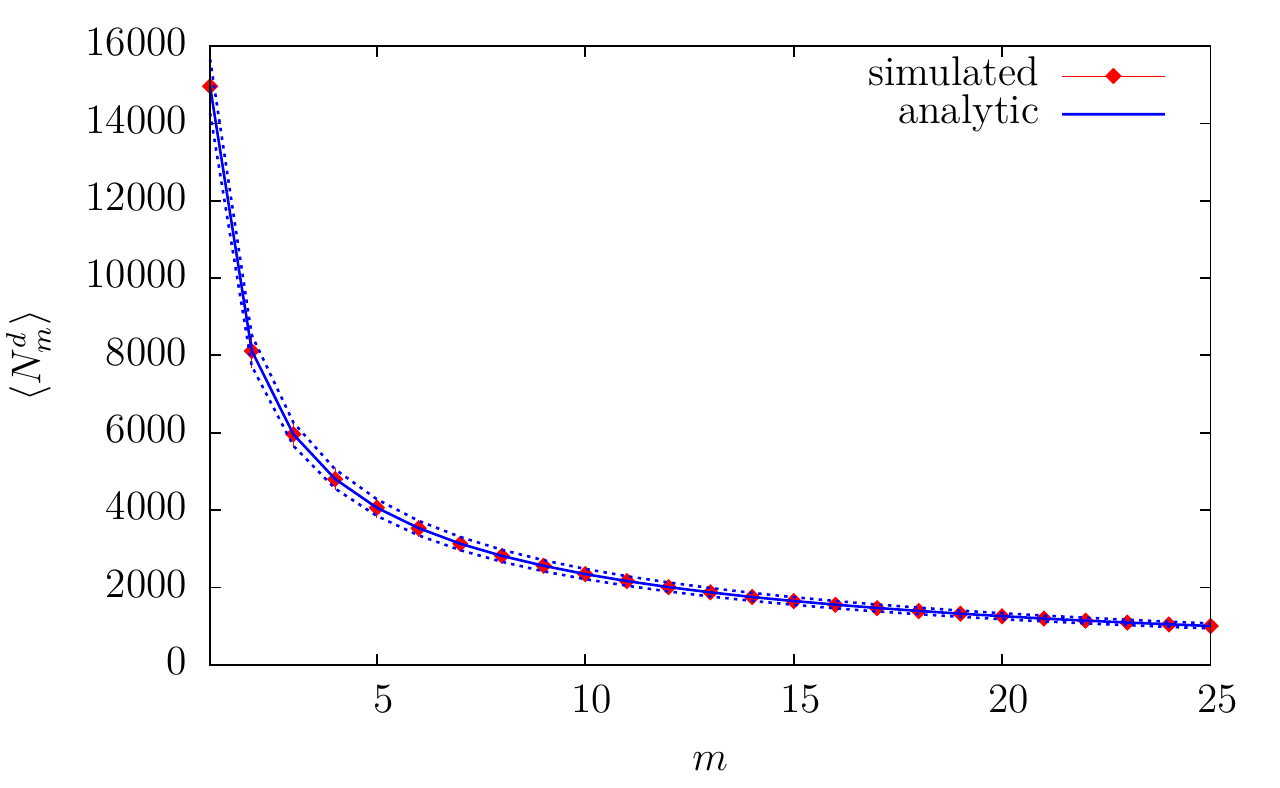}}
\subfloat[4d - 10000 Points]{\includegraphics[width=0.5\textwidth]{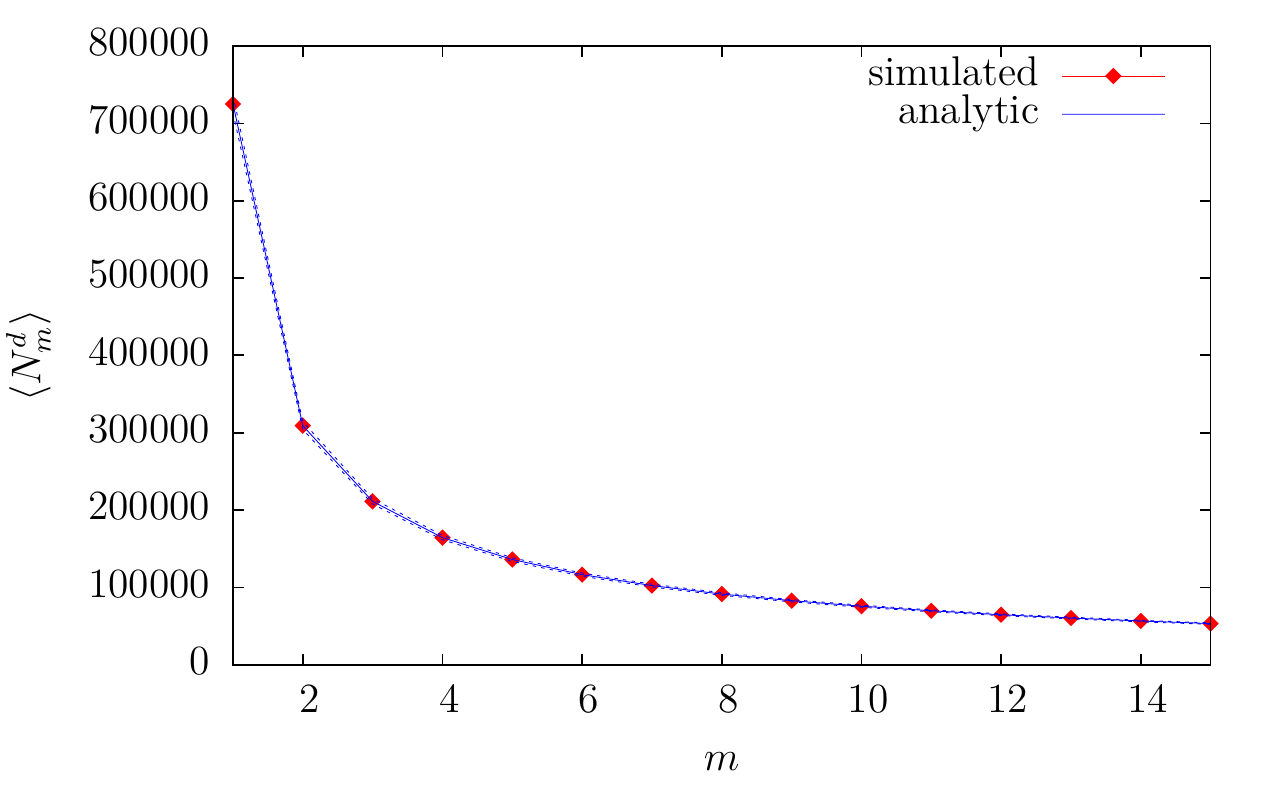}}

\caption{\label{fig:simflat} Simulations of the expectation value of interval abundances
  $\langle\NmCC\rangle$ in flat space for $N=10^{d}$ element causal sets obtained by
  sprinkling  $1000$  times into an interval in flat spacetime. 
  The red dots depict the simulations along with error
  bars.  The solid blue line is $\Nmd(N)$ while the dotted blue lines are $\Nmd(N\pm \sqrt N)$.}
\end{figure}

We first consider the class of causal sets obtained via a Poisson sprinkling into flat spacetime
intervals with $\langle N \rangle = 10^{d}$ elements using the existing Cactus code and calculate
$N_m^d$ for each realisation of a sprinkled causal set, for $d=2,\ldots, 4$.  We consider $1000$
realisations in each case and calculate the standard deviation for the interval abundances. 
We find a remarkable agreement with the analytic curve for $N_m^d$ as
shown in Fig. \ref{fig:simflat}, where we have also plotted the analytic curves for $N\pm \sqrt N$.

\begin{figure}
\centering \includegraphics{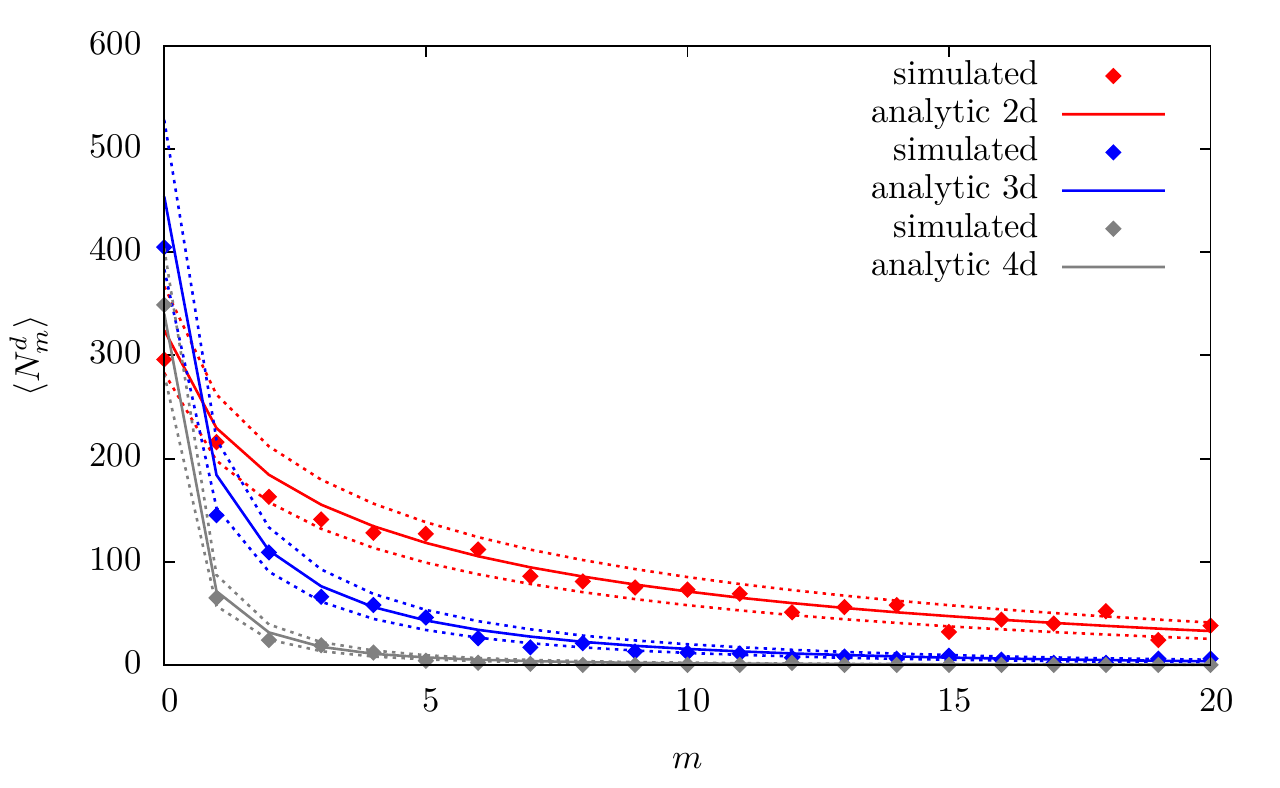}

\caption{\label{fig:singlesprinkle} Simulations in flat space for single realisations of $N=100$
  element causal sets obtained via Poisson sprinkling into flat spacetime intervals. The dots
  represent the  simulations for a single causal set while the solid and dotted lines are  $\Nmd(N)$
  and $\Nmd(N\pm\sqrt N)$, respectively. The agreement is striking.}
\end{figure}

We also find that for single realisations of a sprinkled causal set, the distribution of $N_m^d$
lies well within these curves as shown in \ref{fig:singlesprinkle}.
This plot also shows that the abundance can be used as a continuum dimension estimator.  For causal
sets that are non-manifoldlike this will give a null result since the profile of $\NmCC$ will not
match that of the continuum for any $d$.

\subsection{\label{sub:cut}Examining other topologies}
As the simplest  generalisation of flat spacetime intervals, we consider causal sets that are
sprinkled into flat geometries with non-trivial spatial topology. 
\begin{figure}
	\centering{}
	\includegraphics[width=0.4\textwidth]{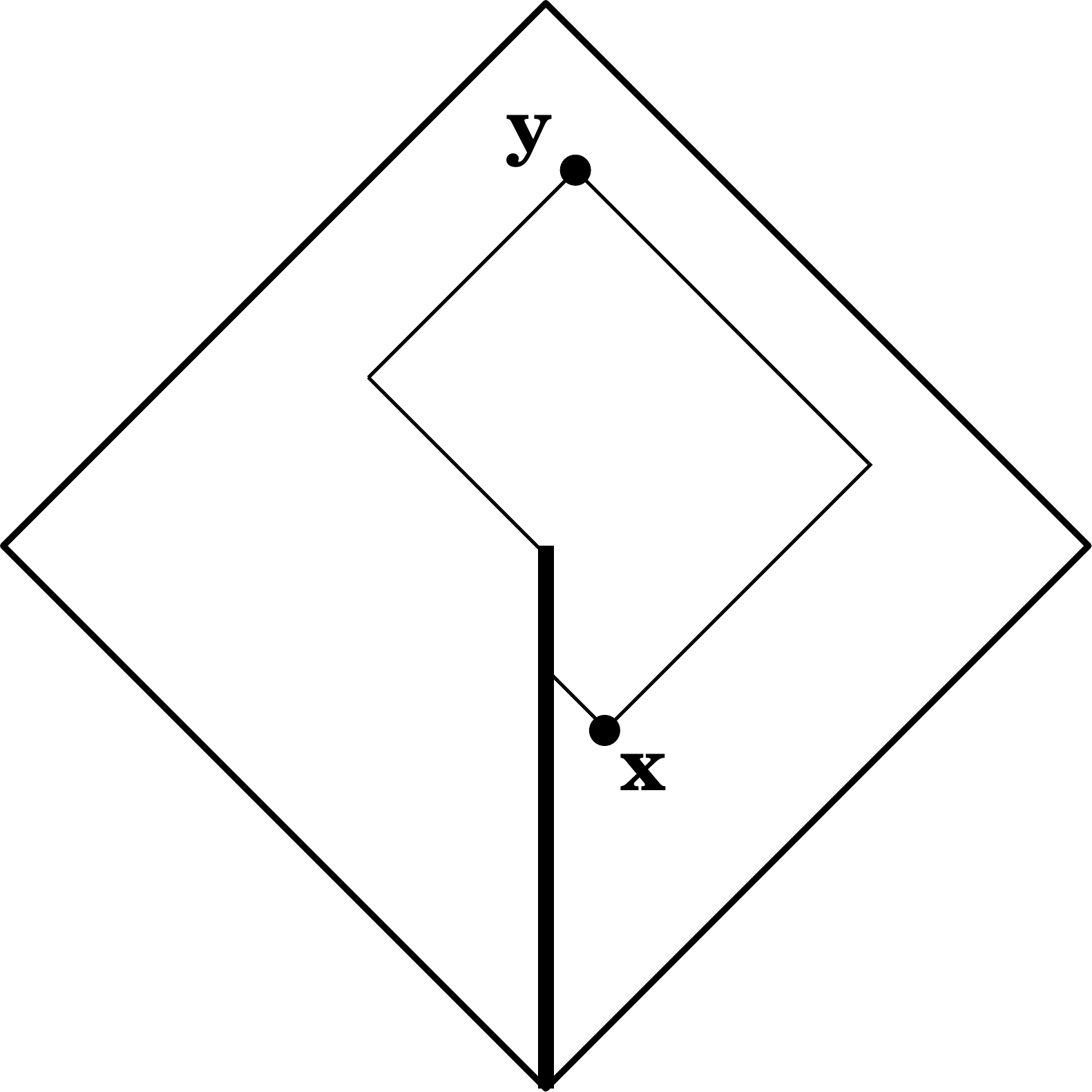}
	\caption{\label{fig:cut} A sketch of the cut-trousers topology.  The Alexandrov
          neighbourhood between the points $x$ and $y$ is modified  by the cut.}
\end{figure}
An example of this is a ``cut-trousers'' topology in $2$-d, with two disjoint spatial intervals
$I\cup I $ joining up to give a single spatial interval $I$, as depicted in Fig. \ref{fig:cut}.

\begin{figure}
\centering
\includegraphics{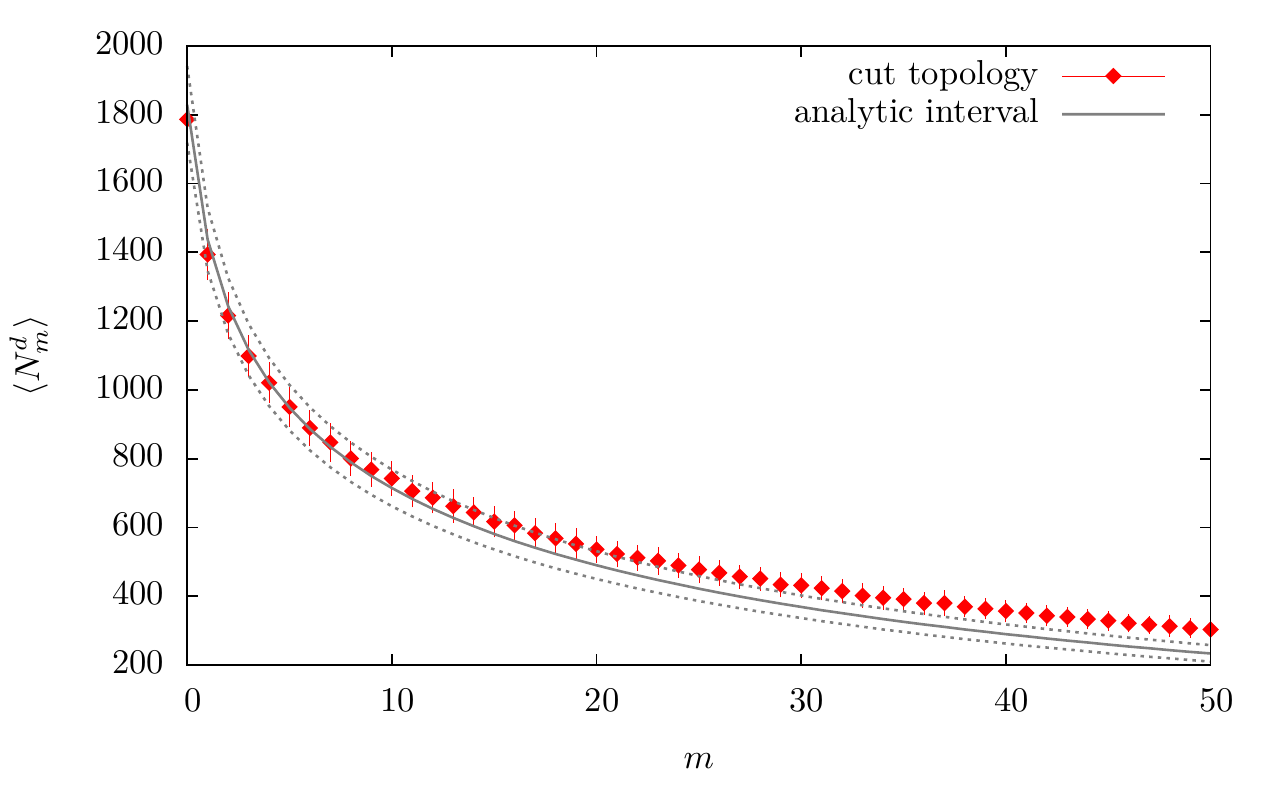}
\caption{\label{fig:simtrousers} $\langle\NmCC\rangle$ for the largest
  intervals contained in  $100$ realisations of $N=1000$ element causal sets obtained from
  sprinklings into the cut-trousers topology.}
\end{figure}

For the plot in Fig. \ref{fig:simtrousers} we obtained $100$ realisations of a $1000$-element causal
set. For each of these, we picked large intervals by choosing a minimal element $p$ and a maximal
element $q$ such that $|I[p,q]|$ is the largest interval. Because of the nature of the topology we
are considering, these intervals are ``incomplete'' if taken to be embedded in flat spacetime, as
shown in Fig. \ref{fig:cut}. We find that this size fluctuates by $380.68 \pm 14.01$ and thus, we can
average over the $100$ realisations to obtain the expectation value of the interval abundances for
$380$ element causal sets.  This is within the standard deviation or fluctuation $\sqrt{380}\approx
19.5$ expected. As shown in Fig. \ref{fig:simtrousers}, the curves $\langle \NmC \rangle$ exhibit a
clear deviation from flat spacetime.

\begin{figure}
  \subfloat[$2$-d $100$ elements]{\includegraphics[width=0.5\textwidth]{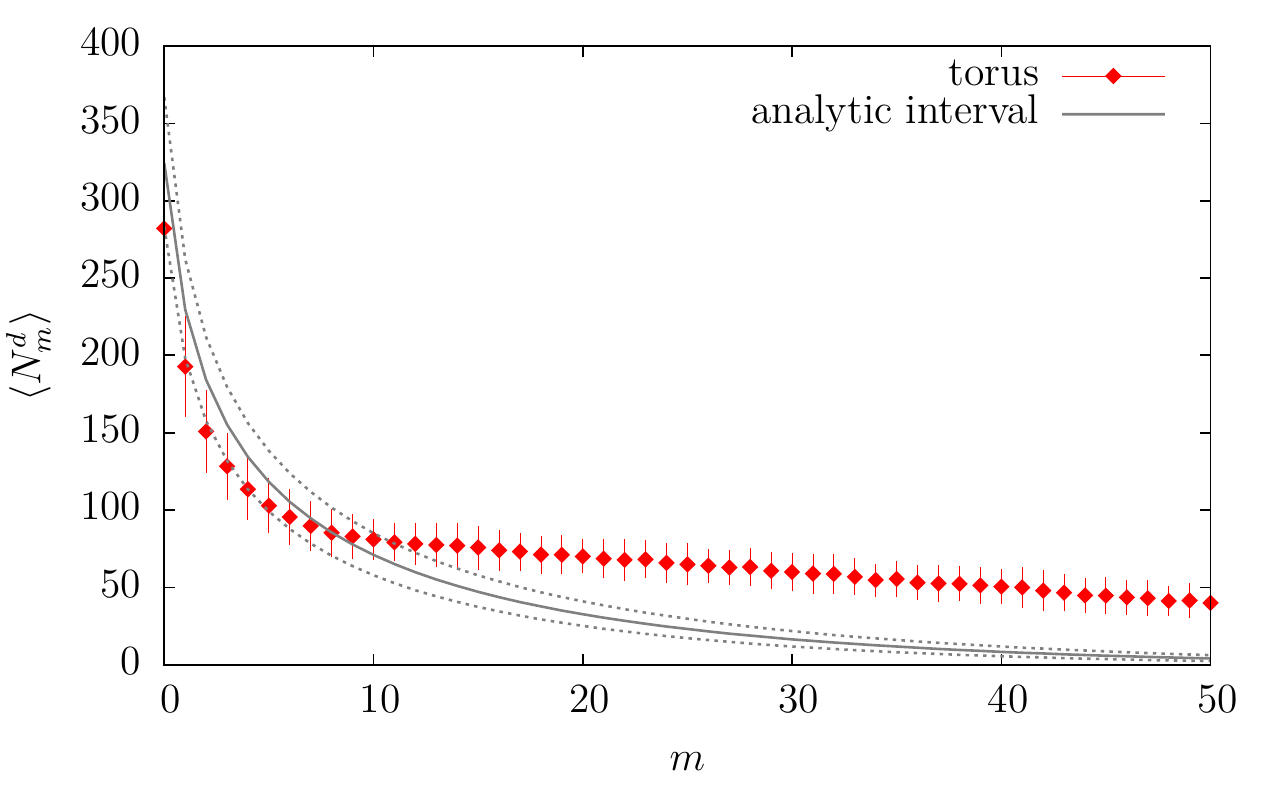}}
  \subfloat[$3$-d $1000$ elements]{\includegraphics[width=0.5\textwidth]{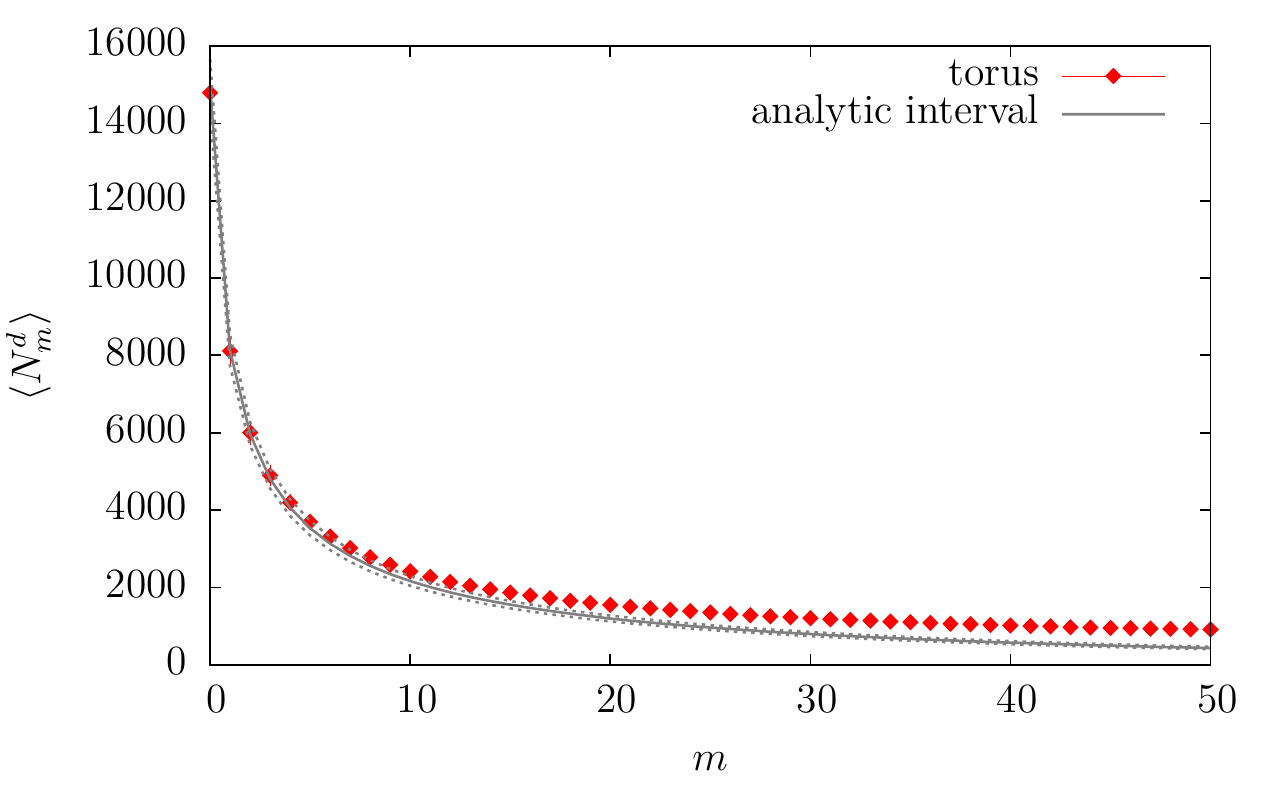}}
  \caption{\label{fig:toruswhole}$\langle\NmCC\rangle$ for  large intervals in causal sets that are
    obtained from $100$ sprinklings into flat spacetimes with  toroidal spatial slices for  $d=2,3$.}
  \end{figure}

We also test causal sets sprinkled into $d=2$ and $d=3$ flat spacetimes with
toroidal spatial topologies, i.e., with $M \equiv \re \times S^1 $ (the cylinder) and $M \equiv \re
\times T^2$, respectively. For the large intervals in $d=2$,
we generate $100$ realisations of $100$-element causal
sets via sprinkling and for $d=3$,  $100$  we generate $100$ realisations of$1000$-element causal sets. 
The results are shown in figure
\ref{fig:toruswhole}.
  
\begin{figure}
\subfloat[$2$-d local and non-local small intervals]{\includegraphics[width=0.5\textwidth]{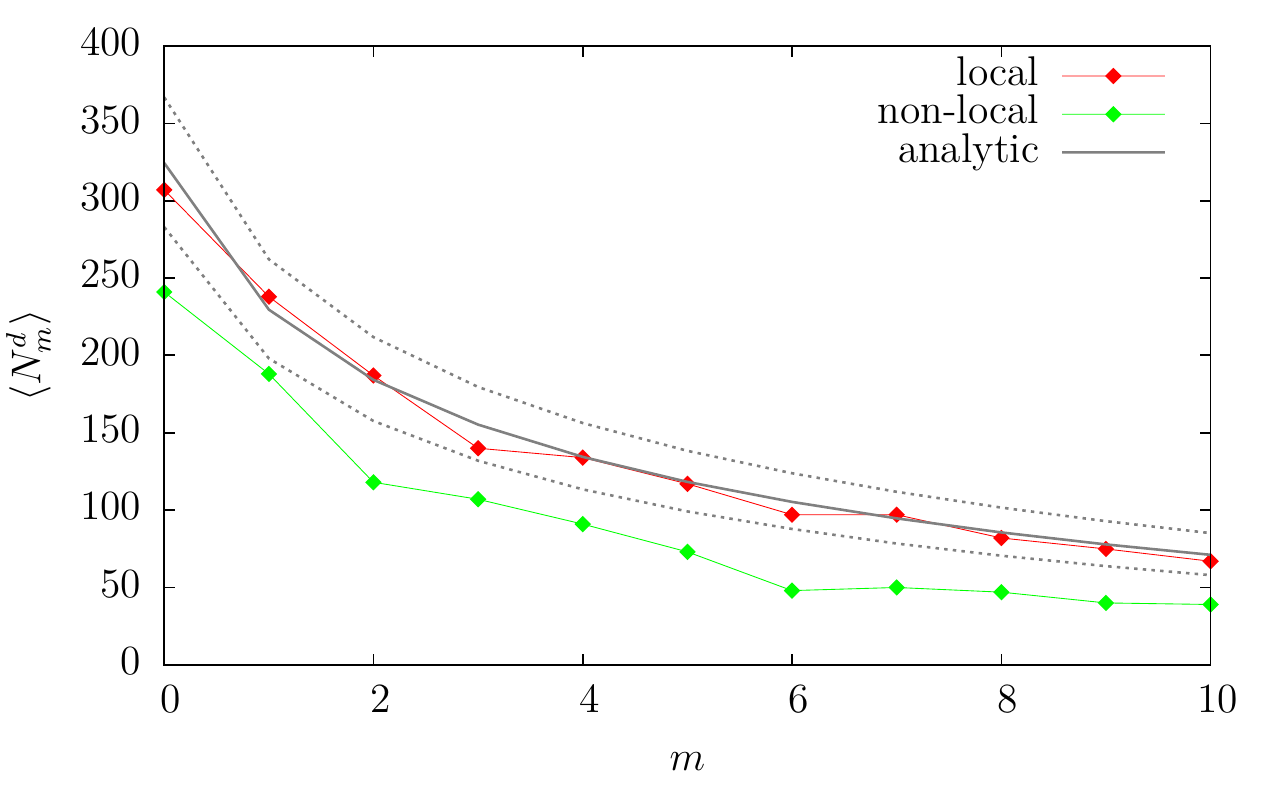}}
\subfloat[$3$-d local and non-local small intervals]{\includegraphics[width=0.5\textwidth]{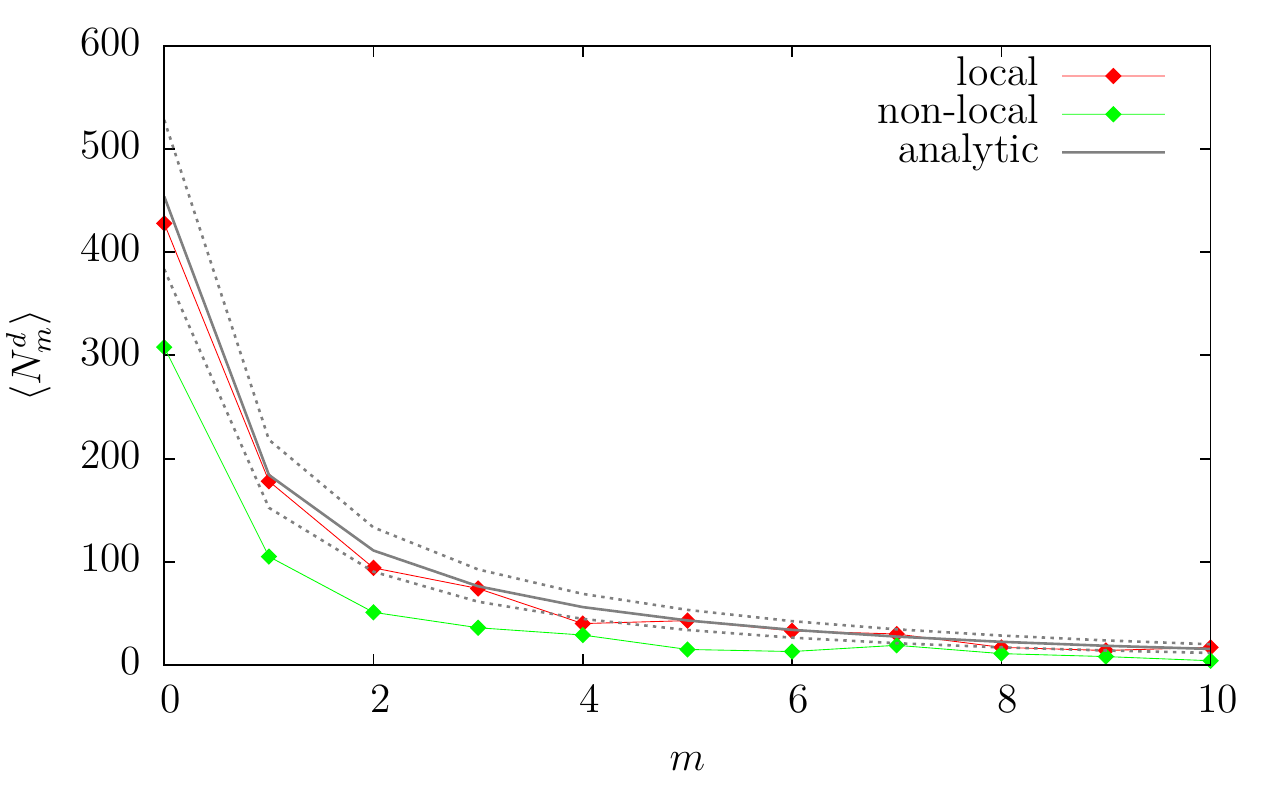}}
\caption{\label{fig:torussmall} $\NmC$ for a small $100$-element  interval causal set $C' $
  contained in a  single $10000$-element causal set obtained by sprinkling into flat spacetime with toroidal spatial slices in $d=2,3$.}
\end{figure}

  For the small intervals we look at single realisations of $10000$-element causal sets and examines
  intervals of size $100$ in both cases. We take more points to obtain a higher density causal set,
  which allows us to find $100$-element intervals that do not probe the topology. As expected, for
  both $d=2$ and $d=3$ the large intervals, which wrap around the compact spatial topology have a
  distribution of intervals which has large deviations from the flat spacetime curve, but most of
  the small intervals do not. That some of the small intervals probe the topology of the space-time
  is due to the non-locality of the causal set. There will always be some small intervals that are
  almost light like, and thus probe the topology of the torus. We illustrate both the ``non-local'' 
  and the ``local'' intervals in Fig. \ref{fig:torussmall}.

\subsection{\label{sub:FRW}Curved spacetime:   FRW and DeSitter Spacetimes}

Next, we  consider causal sets which are sprinkled into  $4$-d  $k=0$ FRW spacetime with metric 
\begin{equation}
\md s^{2} =	-\md t^2 + {a(t)}^2 \left( \sum_{i=1}^4 (dx^i)^2 \right), 
\end{equation}
where 
\begin{align}
	a(t)=&a_{0} \; t^{q}&  \text{with:} &&	q=& \frac{2}{3(1+w)}
\end{align}
with equation of state $p= w \rho$. We show our results from simulations for $w=0,1/3$ and the
deSitter case $w=-1$ (resp. matter, radiation and cosmological constant dominated) as examples. In the deSitter case the deSitter radius arises as a new free parameter, we chose a radius of $\ell =1.3$. For each of these we examine  $100$  realisations of $\langle N
\rangle=1000$-element causal sets and find the average interval abundances. We expect
significant deviations from the flat spacetime case, due to the non-trivial spacetime curvature
and our simulations do confirm this expectation.

\begin{figure}
\centering
\includegraphics{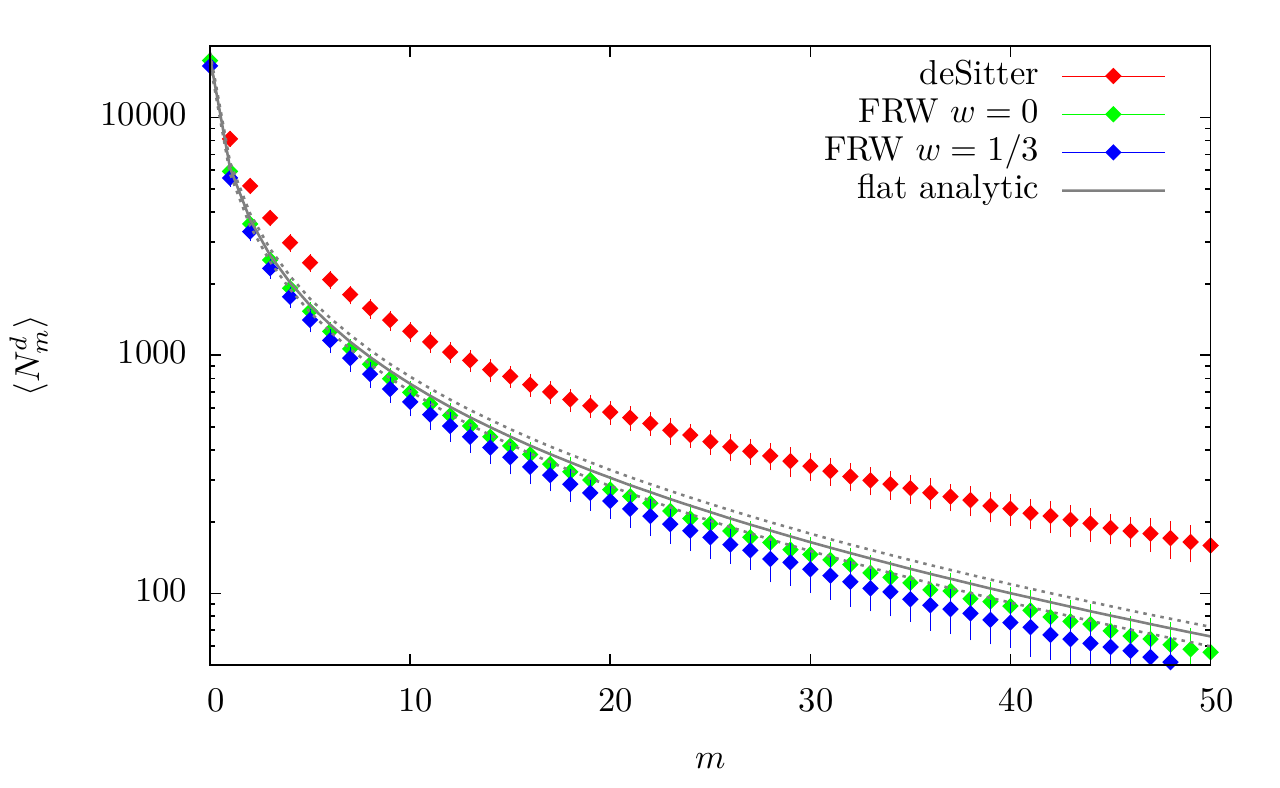}
	\caption{\label{fig:FRW_whole} $\langle \NmCC \rangle$ for   $N=1000$ element causal sets obtained from
          sprinkling $100$ times into $4d$ FRW spacetimes which are $\Lambda$, matter or radiation dominated.}
\end{figure}

We first show that the large intervals do not follow the flat spacetime
  characteristic curve. As for the intervals with non-trivial topology, the size of the intervals varies
  as $N_x \pm \sqrt N_x$. The results for all three choices clearly show the effect of curvature on
  the interval abundances.

\begin{figure}

\subfloat[FRW with $w=0$]{\includegraphics[width=0.58\textwidth]{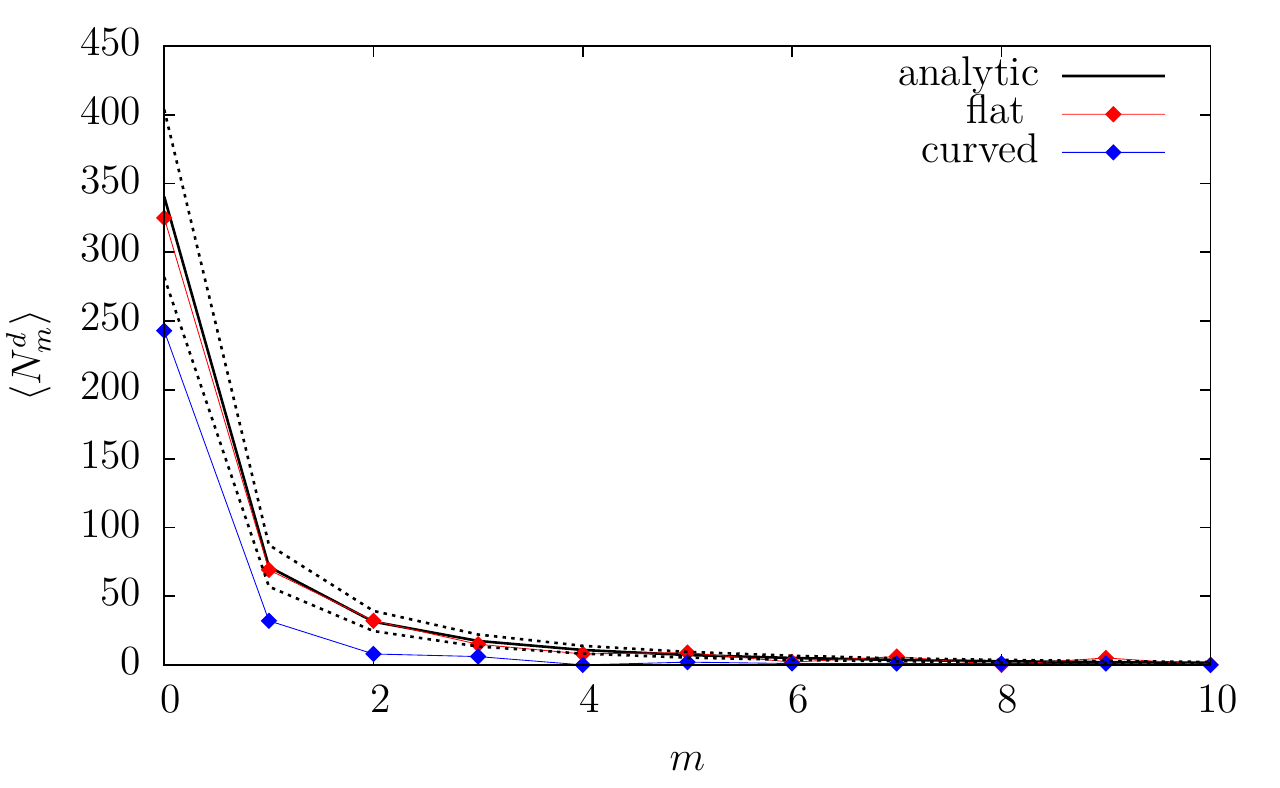}  \hspace{15pt} \includegraphics[width=0.33\textwidth]{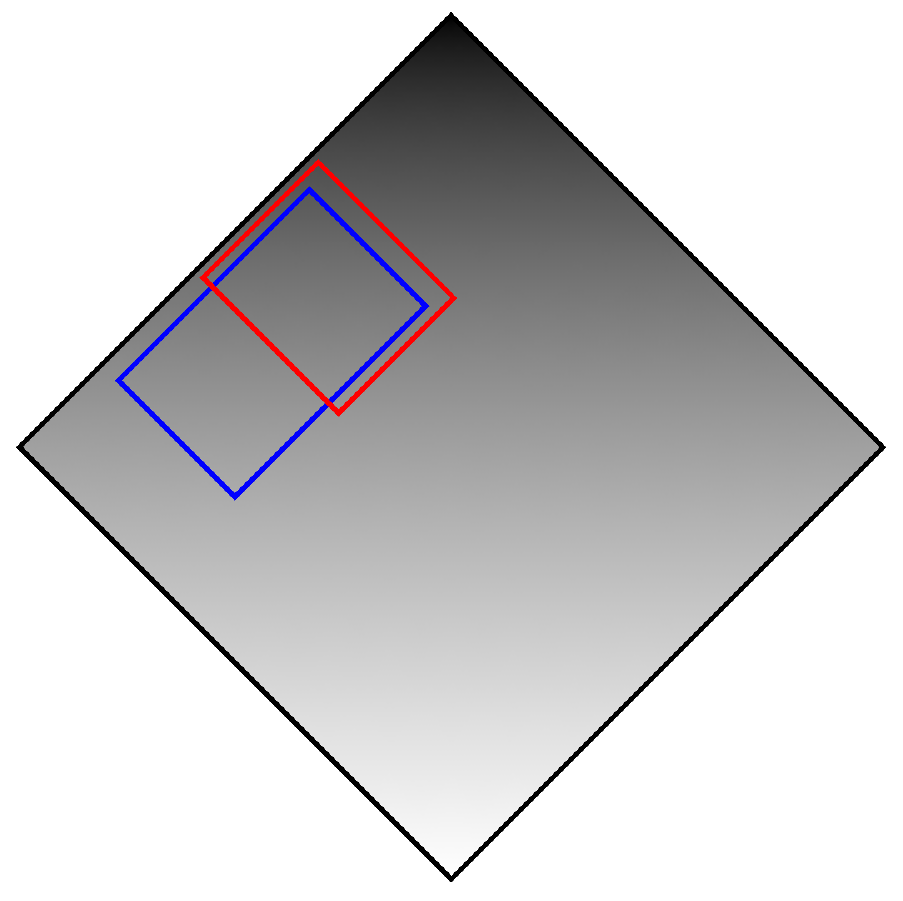}}\\
\subfloat[FRW with $w=\frac{1}{3}$]{\includegraphics[width=0.58\textwidth]{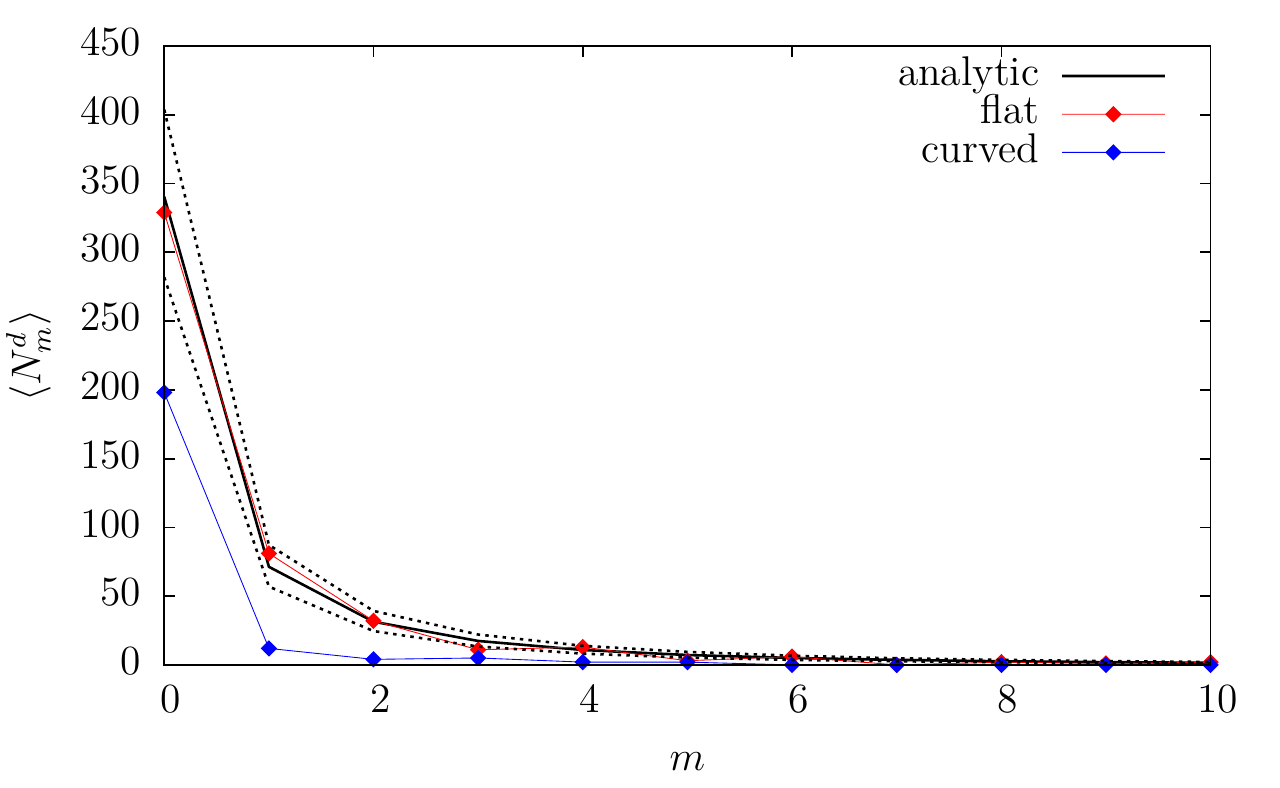} \hspace{15pt} \includegraphics[width=0.33\textwidth]{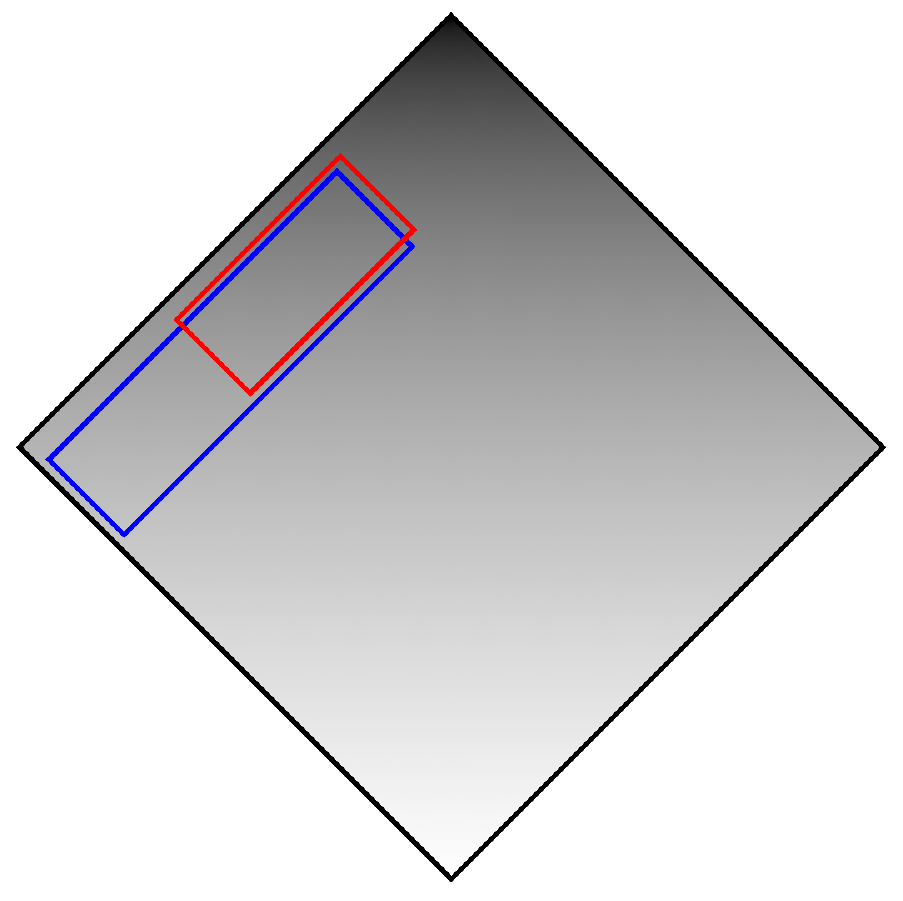}}\\

\caption{\label{fig:FRW_subsets} Single realisations of small interval causal sets $C'$ contained in an
  $N=10000$ element causal set $C$ obtained from a sprinkling into $4d$ FRW spacetimes which matter or radiation dominated. The sketches on the
  right hand side show which intervals are local and which non-local, while  the shading indicates
  the scale factor of the universe.} 
\end{figure}

\begin{figure}
\subfloat[$100$ element deSitter intervals]{\includegraphics[width=0.48 \textwidth]{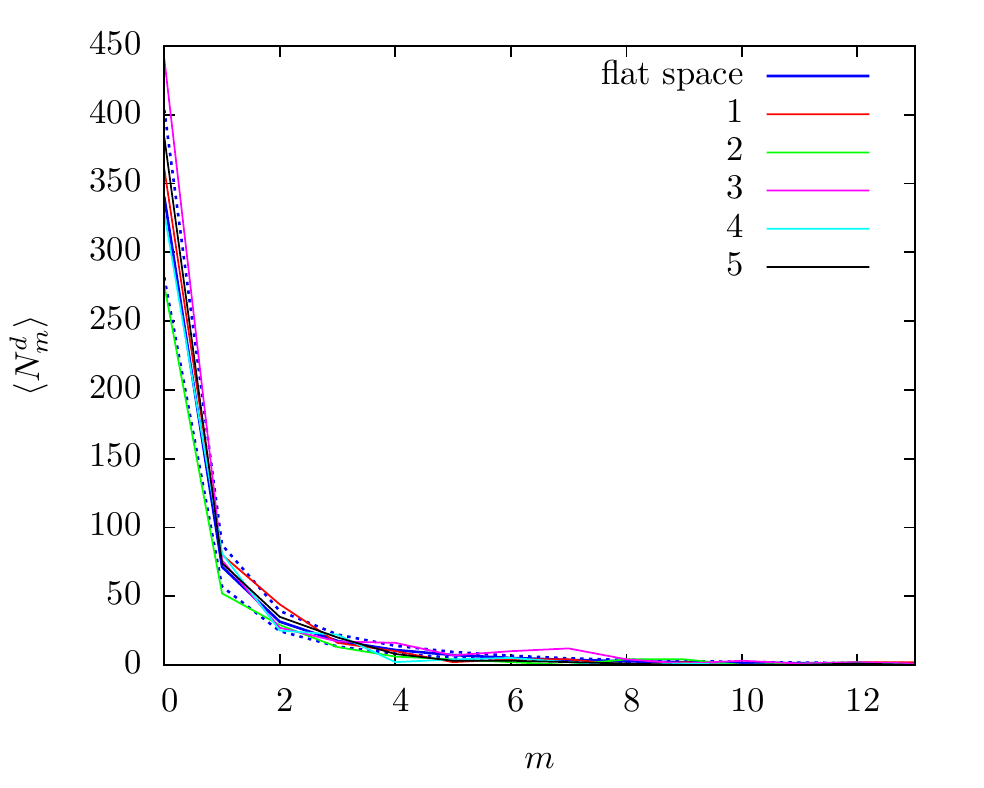}}
\subfloat[$2000$ element deSitter intervals]{\includegraphics[width=0.48 \textwidth]{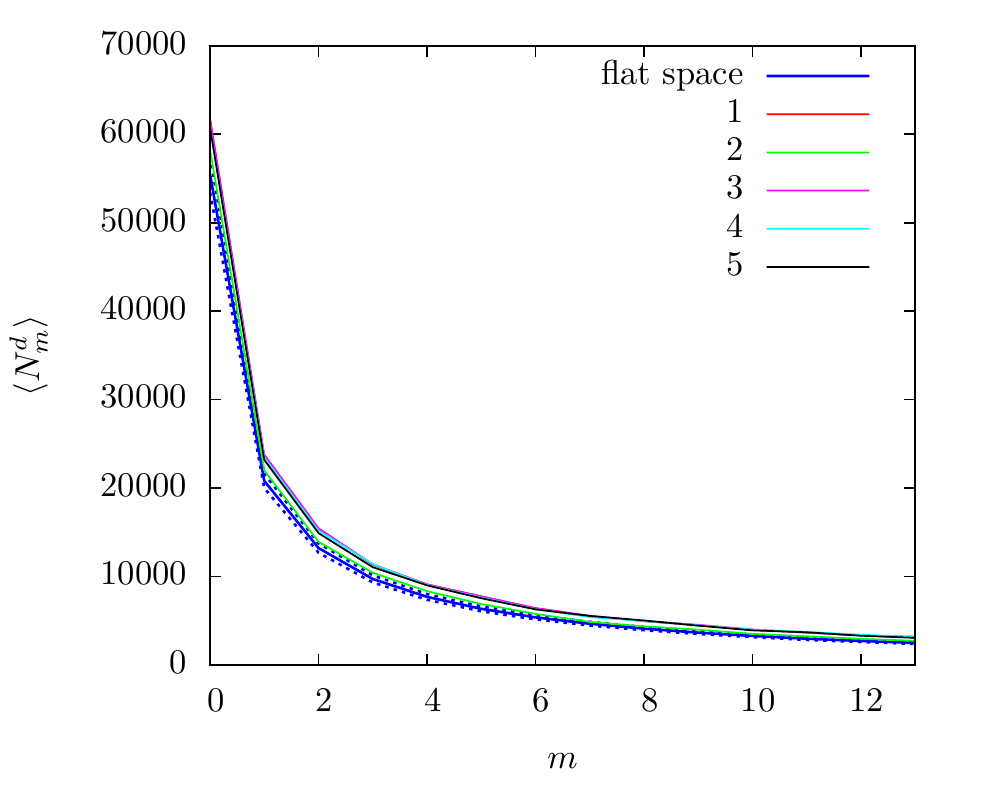}}
\caption{\label{fig:deSitter} 5 single realisations of small interval causal sets $C'$ contained in a $N=10000$ element causal set $C$ obtained from a sprinkling into $4d$ deSitter spacetime.}
\end{figure}

To test manifoldlikeness of the causal set, we examine the intervals of size $100$ containing a
randomly chosen element in a single realisation of an $10000$ element causal set. For $w=\frac{1}{3}$ and $w=0$
we found that there are intervals for which the abundances follow the flat spacetime curve and
those which demonstrate significant deviations.  As in the case of non-trivial topology, these
latter interval neighbourhoods must sample a region in which the scale of flatness $\curv^{-1}$ is
small, i.e., they are elongated intervals.  The result is shown in Fig. \ref{fig:FRW_subsets}. The
coloured boxes indicate the plotted intervals in co-moving coordinates, while the shading indicates
the scale factor $a(t)$ which needs to be taken into consideration when comparing the
intervals. Although the elongated intervals do not fit the curve for flat $4$d space, they are still
not in agreement with higher or lower dimensional spacetime.  If the same type of test is done on
flat sprinkled causal sets, sprinkled sufficiently densely, there are no such stark differences
between elongated and flat intervals. The same is true for intervals in deSitter spacetime.
In figure \eqref{fig:deSitter} we show intervals of size $100$ and $2000$ picked out of a $10000$ element deSitter sprinkling.
They are as similar to each other as would be the case for flat space. This is because deSitter space is maximally symmetric.
While the $100$ intervals are in agreement with flat $4$d spacetime the $2000$ element sets all show a significant deviation from flatness.

\subsection{Causal sets obtained from Transitive percolation}
Causal sets should ideally not be created by sprinkling but grow naturally from some form of
process. One such process, put forward by Rideout and Sorkin is transitive percolation
\cite{Rideout:1999ub,Ahmed:2009qm}.

In transitive percolation the causal set grows iteratively, one element at a time. Each element gets
added and then connected to the older elements with a certain probability.  The probability for an
element at step $n$ to be connected to an element at step $n-1$ is denoted as $p$, which is the only
free parameter. In \cite{Ahmed:2009qm,Rideout:1999ub} causal sets of this type have been observed to have some manifoldlike characteristics. One thing that was examined was the functional relation between the proper time distance of two points and the volume that lies causally between them. It was found that for a variety of parameter combinations this curve can be well fit with the corresponding volume of a deSitter spacetime, using the deSitter radius $\ell$ and a proportionality factor between the length of the longest chain and the proper time $\tau$ as free parameters.
To examine if percolated causal sets also appear manifoldlike under our new test we picked some of the possible parameter combinations, summarised in table \ref{tab:params}. 

\begin{table}
\caption{\label{tab:params}Three sets of parameter values from \cite{Ahmed:2009qm} which we have examined.}
\centering
\begin{tabularx}{\textwidth}{X X X X X}
\toprule
p & N & d & $\ell$ & m  \\
\midrule
0.03 & 1000 & 3d & $2.331 \pm 0.011$ & $1.046 \pm 0.006$	\\
0.01 & 2000 & 3d & $4.086 \pm 0.028$ & $1.136 \pm 0.006$	\\
0.005 & 15000 & 4d & $ 6.20 \pm 0.12$& $1.710 \pm 0.013$	\\
\bottomrule
\end{tabularx}

\end{table}
\begin{figure}
{\centering
\subfloat[ p=$0.005$ N=$15000$, average size $\langle N \rangle=1259.1 \pm  47.8$]{\includegraphics{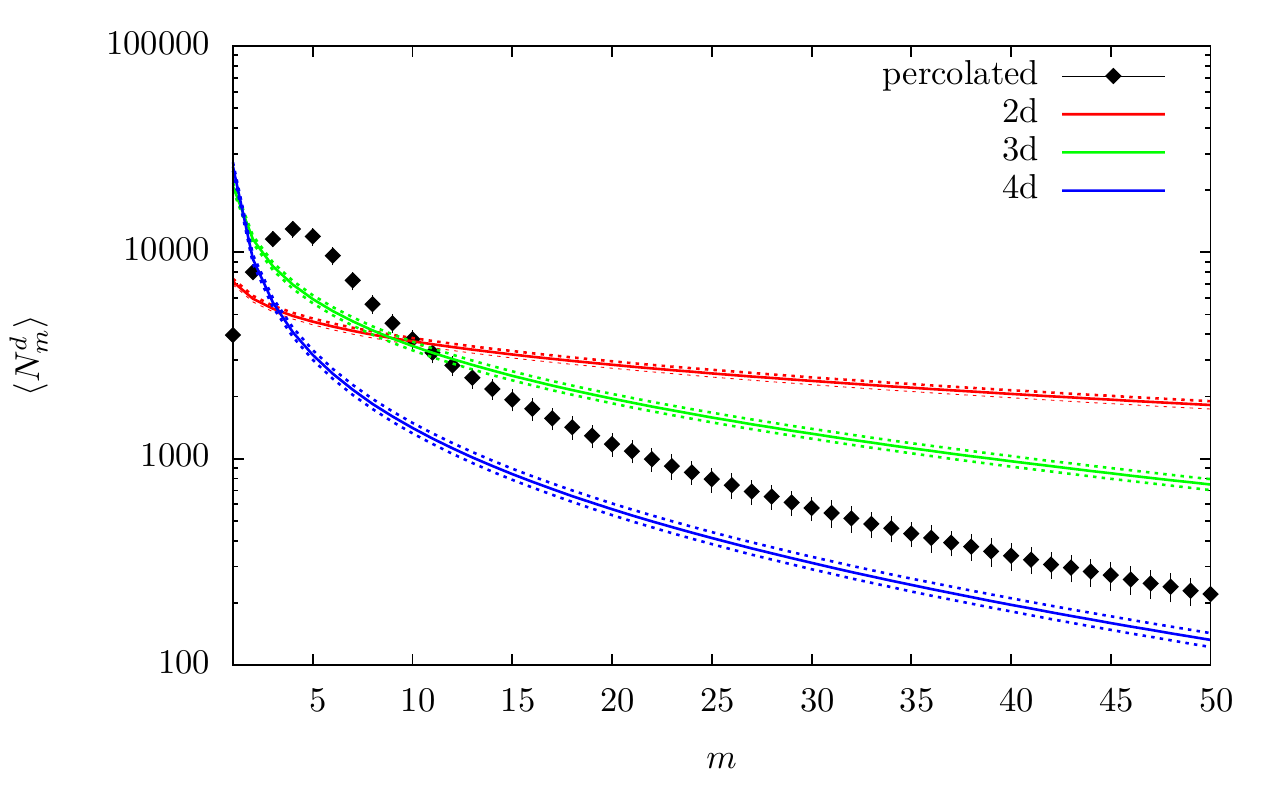}}

}

\subfloat[ p=$0.03$ N=$1000$, average size $\langle N \rangle=335.2 \pm 25.9$ ]{\includegraphics[width=0.5\textwidth]{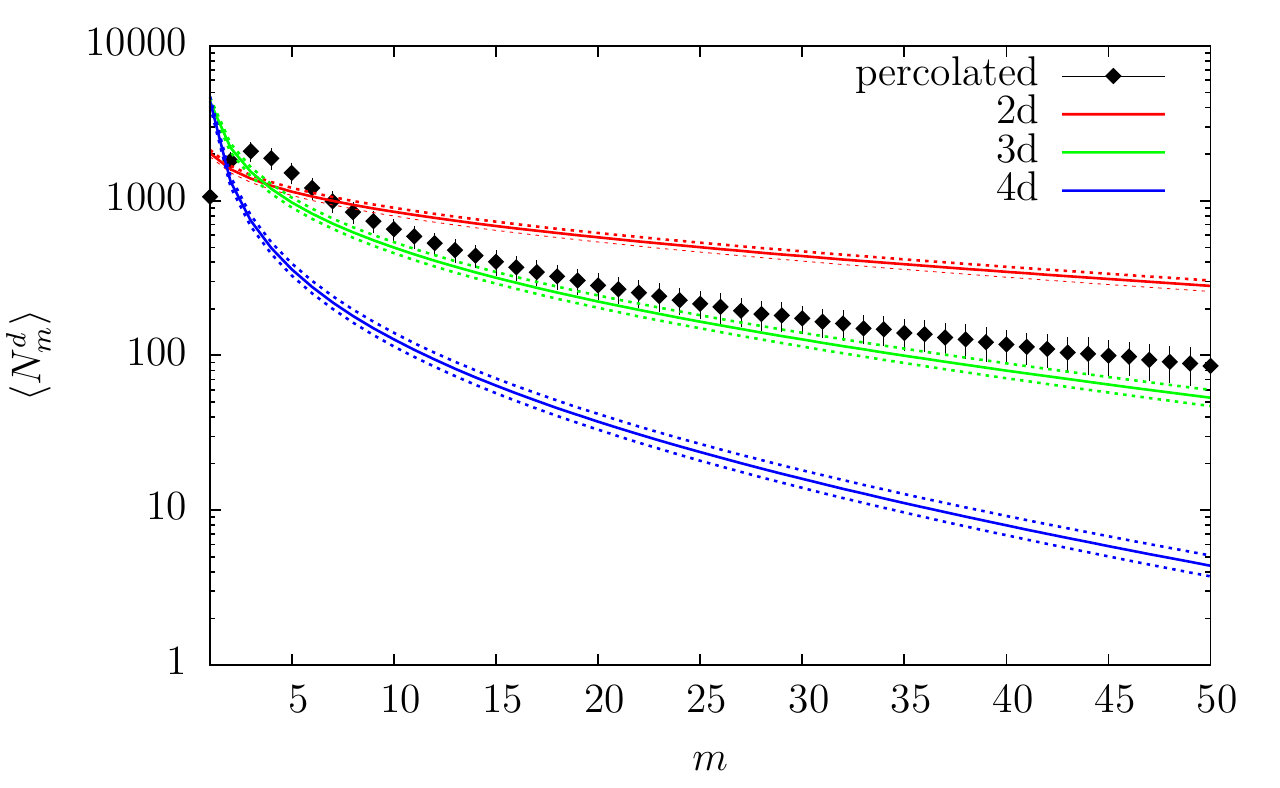}}
\subfloat[ p=$0.01$ N=$2000$, average size $\langle N \rangle=695.7 \pm 53.5$]{\includegraphics[width=0.5\textwidth]{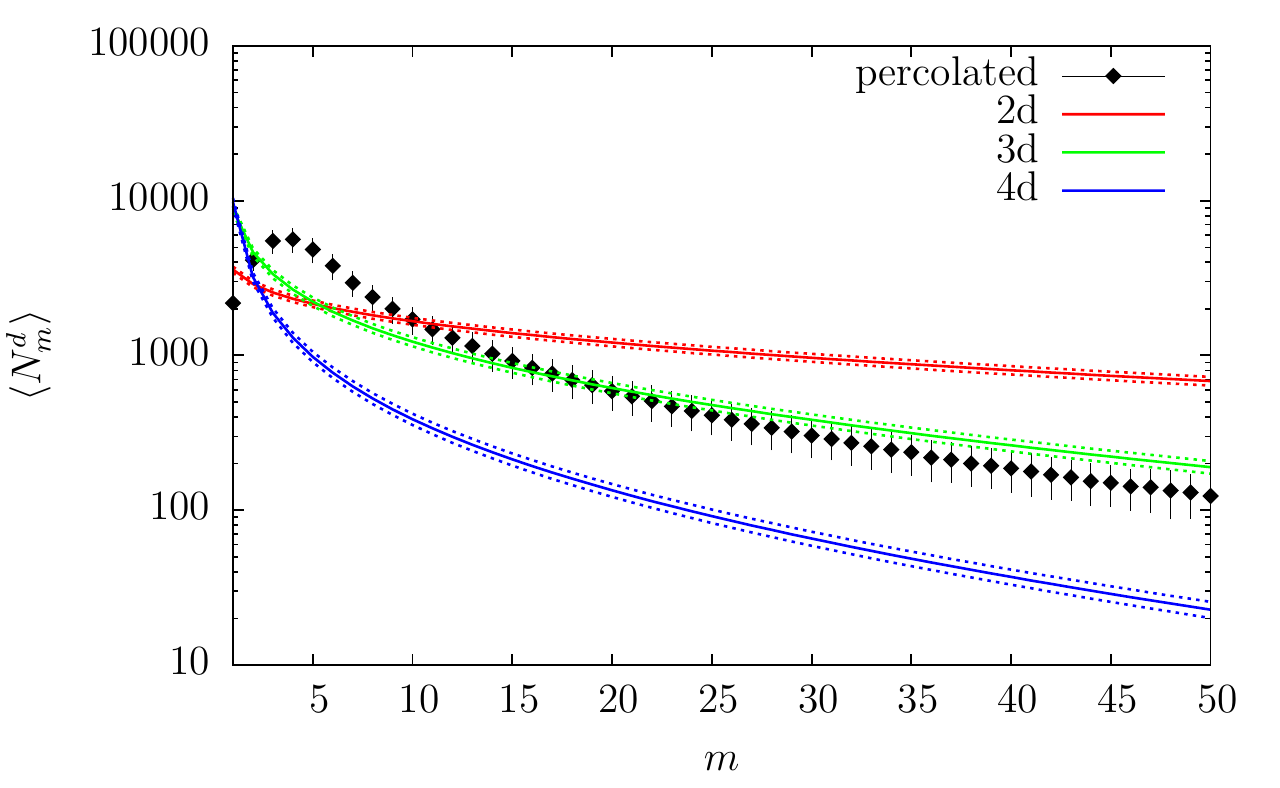}}

\caption{\label{fig:percolation}The $\langle \NmCC \rangle $  for percolated causal sets
 whose longest chain has $20$ elements,
are shown  in black and compared with the $\Nmd$ for $d=2,3,4$.  }
\end{figure}
In the paper they fit the curve to only those intervals which for a given proper time, had the largest volume.
We followed this up in finding those intervals and measuring their interval abundances. 

We created $100$ percolated causal sets for each of the parameter combinations stated in table
\ref{tab:params} and calculated the average interval abundances for intervals of different proper
times. In Fig. \ref{fig:percolation} we plot this for the intervals of height $20$. The behaviour
of the interval abundance is similar for heights between around $10-50$ which is roughly the range
of heights to which the deSitter volume profile was fitted in \cite{Ahmed:2009qm}. Indeed there is a
striking dissimilarity with flat spacetime: the abundances $\NmC$ show a maximum at some $m >0$,
unlike in flat spacetime for which $\NmC$ is a maximum for the links, i.e., $m=0$.  This maximum
shifts to larger $m$ as one examines larger Alexandrov intervals. One could perhaps argue
  that the different shape of the intervals could arise  from curvature. However, the
  difference in  shape persists even for very small intervals, which should look flat.This
  gives a strong indication that percolated causal sets are not manifoldlike.

Interestingly,  while the interval abundance is clearly not that of flat spacetime it does converge
towards the interval abundance of the dimension measured in  \cite{Ahmed:2009qm} for large
intervals.  While the abundance of links and 1-element  intervals for the percolated causal sets  is very
different from the analytic prediction, it  falls off  monotonically after the maximum and appears
to get closer to the analytic prediction for manifoldlikeness. 

\begin{figure}
{\centering
\subfloat[ p=$0.005$ N=$15000$, average size $\langle N \rangle=1185.8 \pm  23.16$]{\includegraphics{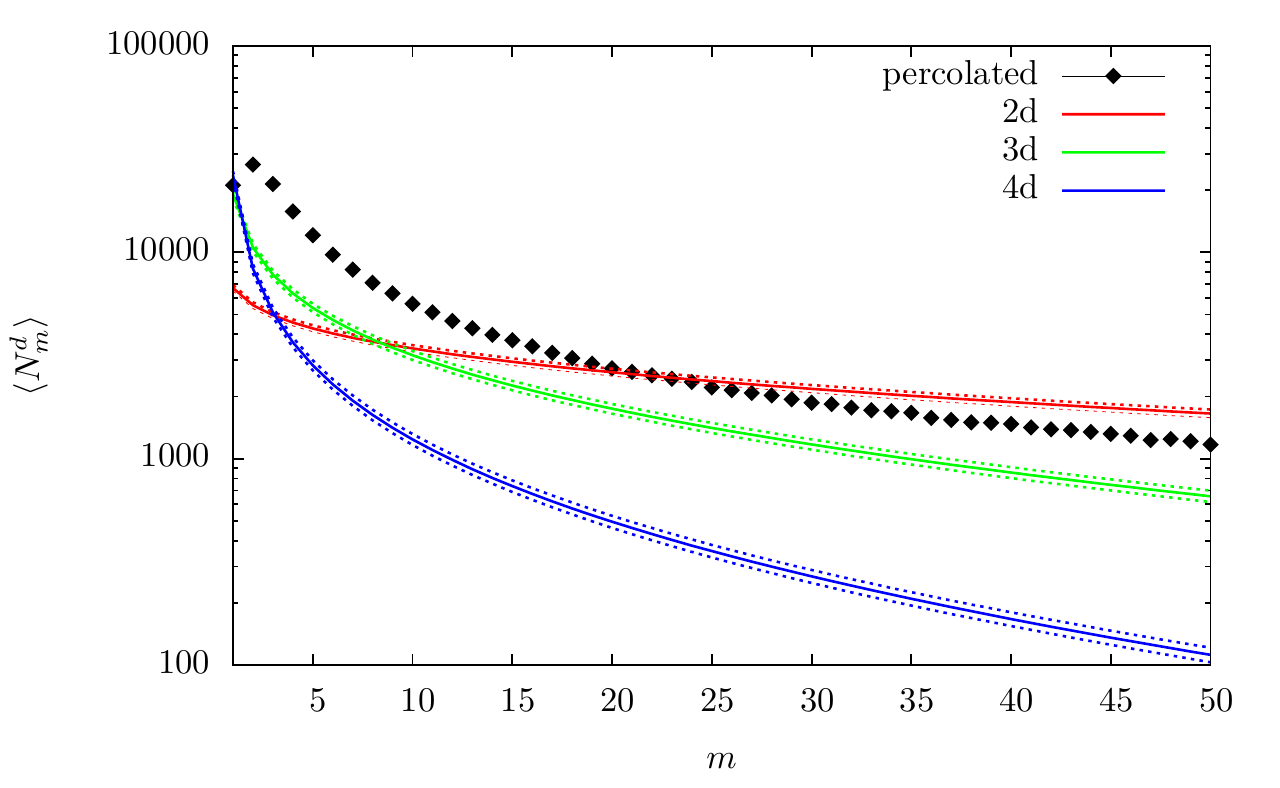}}

}

\subfloat[ p=$0.03$ N=$1000$, average size $\langle N \rangle=249.5 \pm 11.7$]{\includegraphics[width=0.5\textwidth]{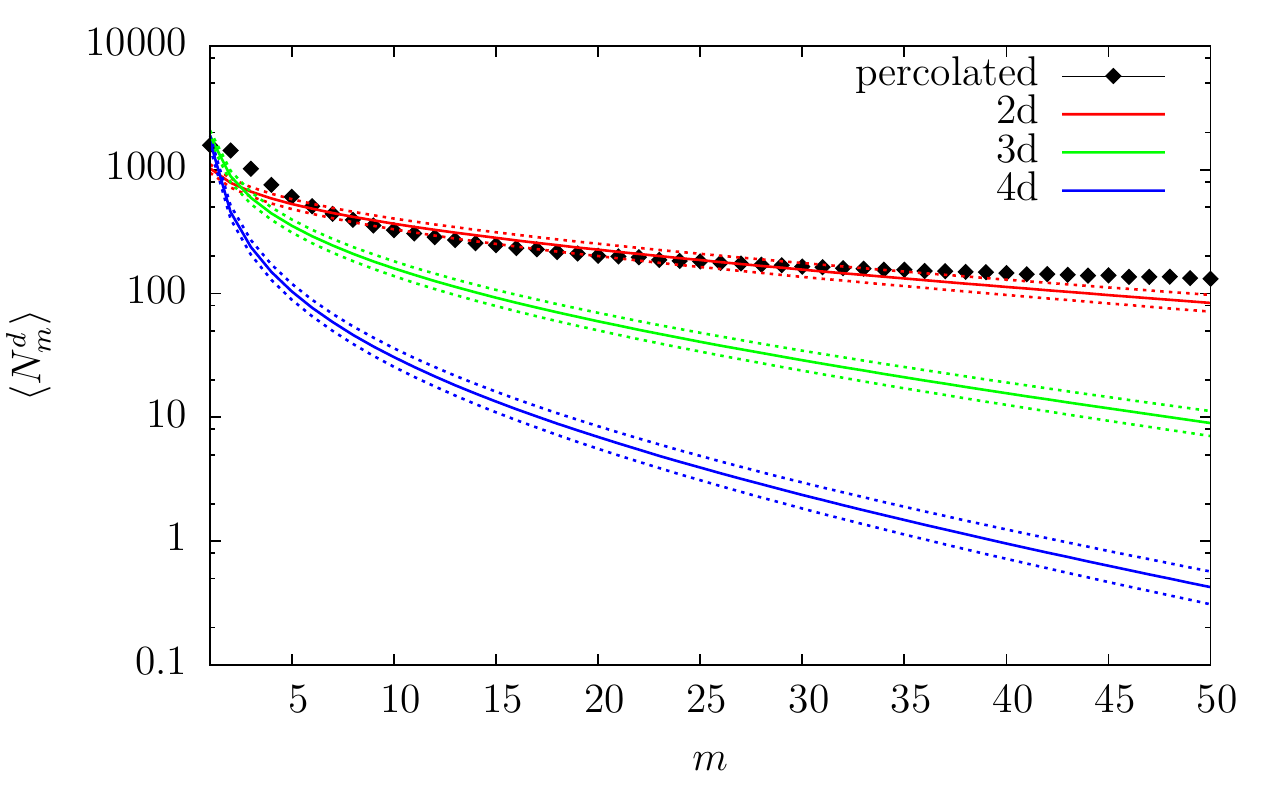}}
\subfloat[ p=$0.01$ N=$2000$, average size $\langle N \rangle=615.5 \pm 30.3$]{\includegraphics[width=0.5\textwidth]{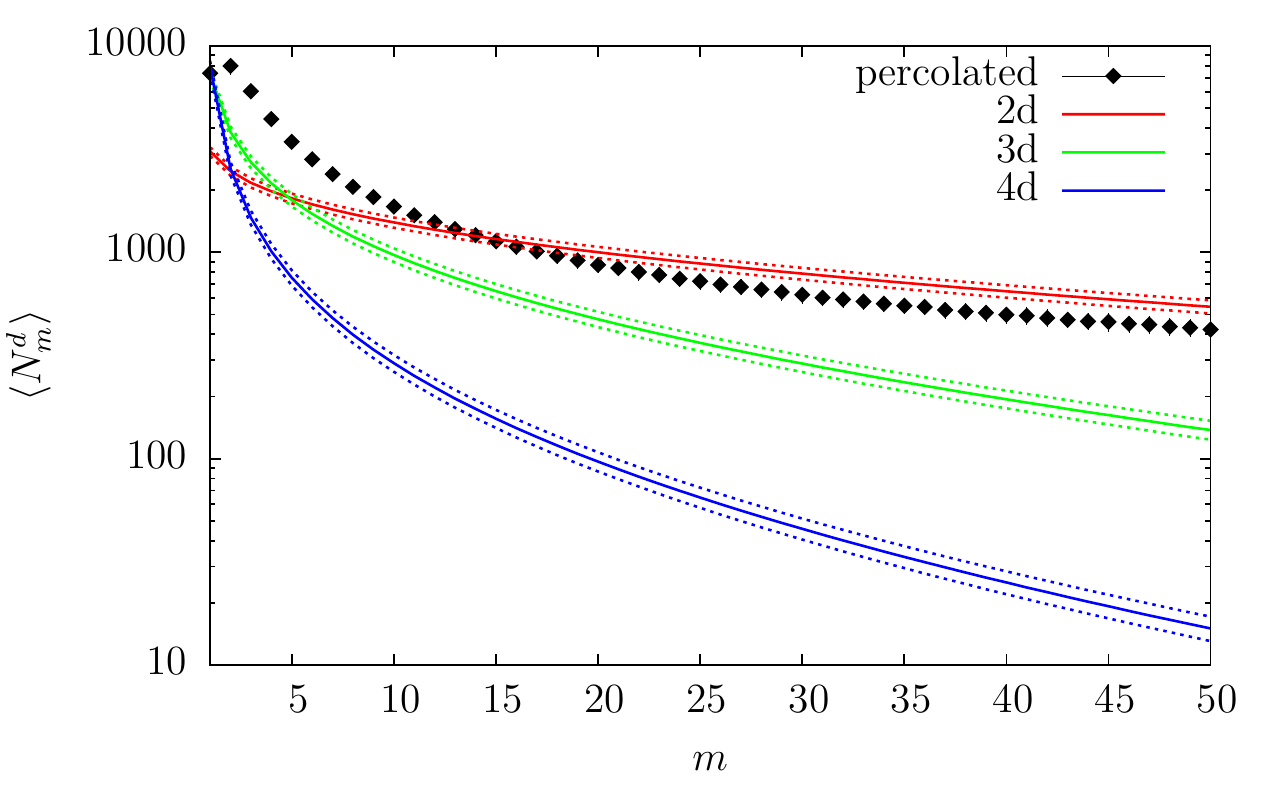}}

\caption{\label{fig:coarsegrained}The $\langle \NmCC \rangle$ for {coarse grained} percolated causal
  sets, whose longest chain is $20$ elements long, is shown in  black and compared with the $\Nmd$  for $d=2,3,4$. }
\end{figure}

This suggests that perhaps the percolated causal sets, while not manifoldlike in the small, might
be manifoldlike at a coarse grained level and hence satisfy our test.  The coarse graining procedure
involves keeping each element of $C$ with a certain probability $P$ . 
In Fig. \ref{fig:coarsegrained} we show the interval abundance for coarse grained transitive
percolated causal sets. We choose $P=0.25$, used the same values of $p$ as before, and fixed the
number of elements such that it would agree with table \ref{tab:params} after coarse graining. 
Fig. \ref{fig:coarsegrained} shows that coarse graining does change the structure of the interval
abundances significantly, since the maximal abundance moves to smaller $m$
and the abundances become a monotonically decreasing function of $m$. However, despite this promising behaviour, the detailed curve differs strongly from the
$\Nmd$. Of course the size of the parameter space makes it hard to make a stronger claim, but for
coarse-grainings where three-quarters, half, one quarter or one tenth of the points were kept we did
not find agreement with the analytic curve for flat spacetime.  Further study to compare to the
interval abundance for deSitter space might be useful, but first attempts at it do not indicate a
substantial change in the results.

\subsection{\label{sub:nonmanifold} Non manifoldlike causal sets}
\begin{figure}
\centering
\includegraphics[width=0.5\textwidth]{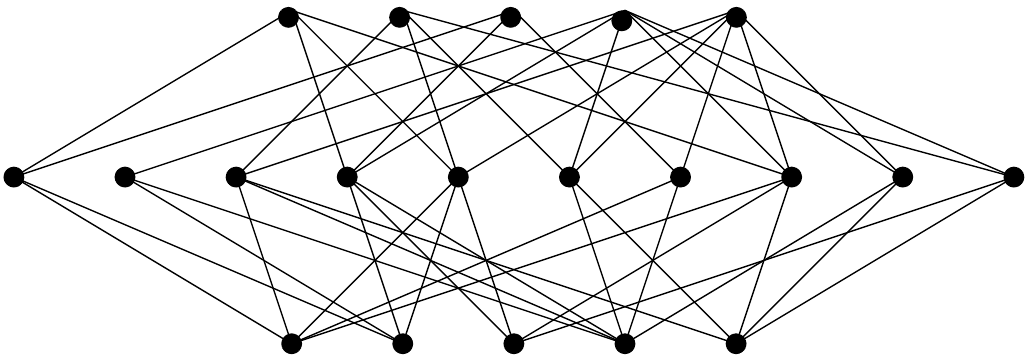}
\caption{\label{fig:KR} An example of a small KR- order}
\end{figure}

\begin{figure}
\large
\subfloat[chain]{\includegraphics[width=0.5\textwidth]{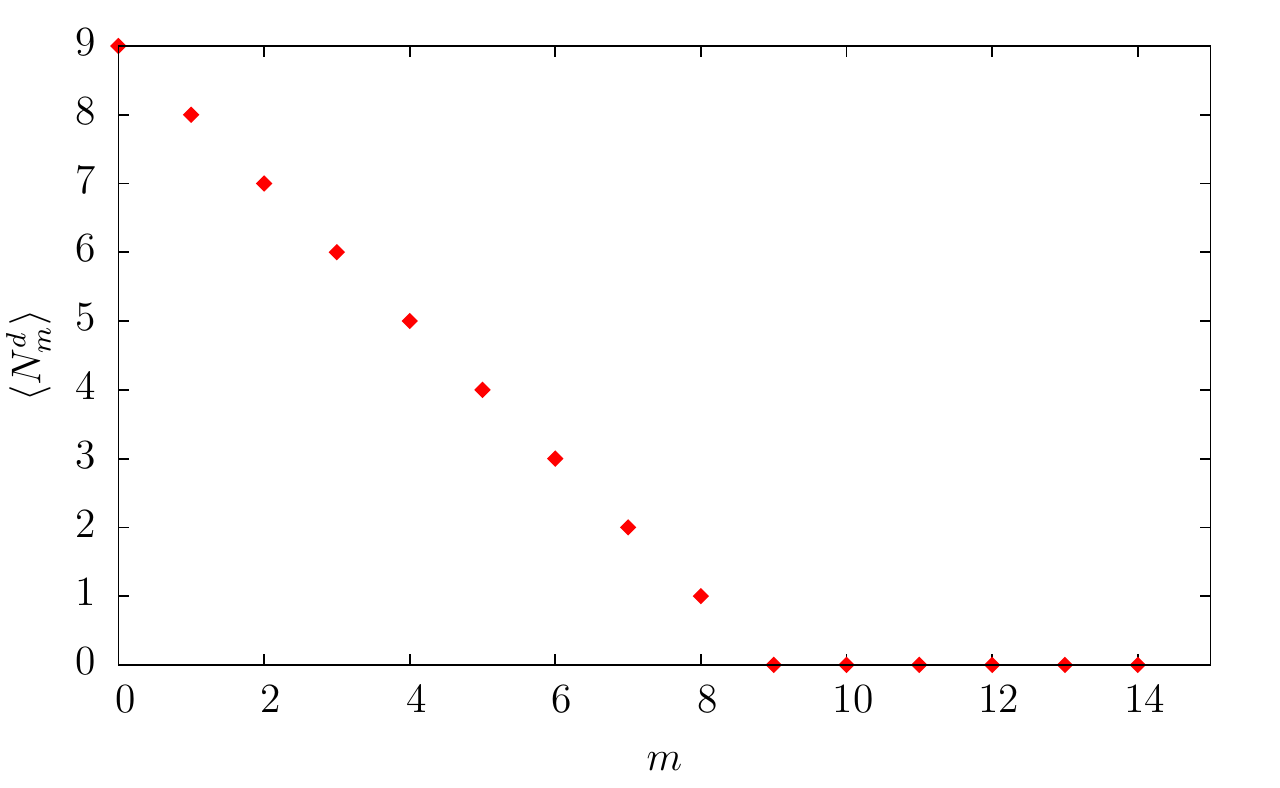}}
\subfloat[KR-order]{\includegraphics[width=0.5\textwidth]{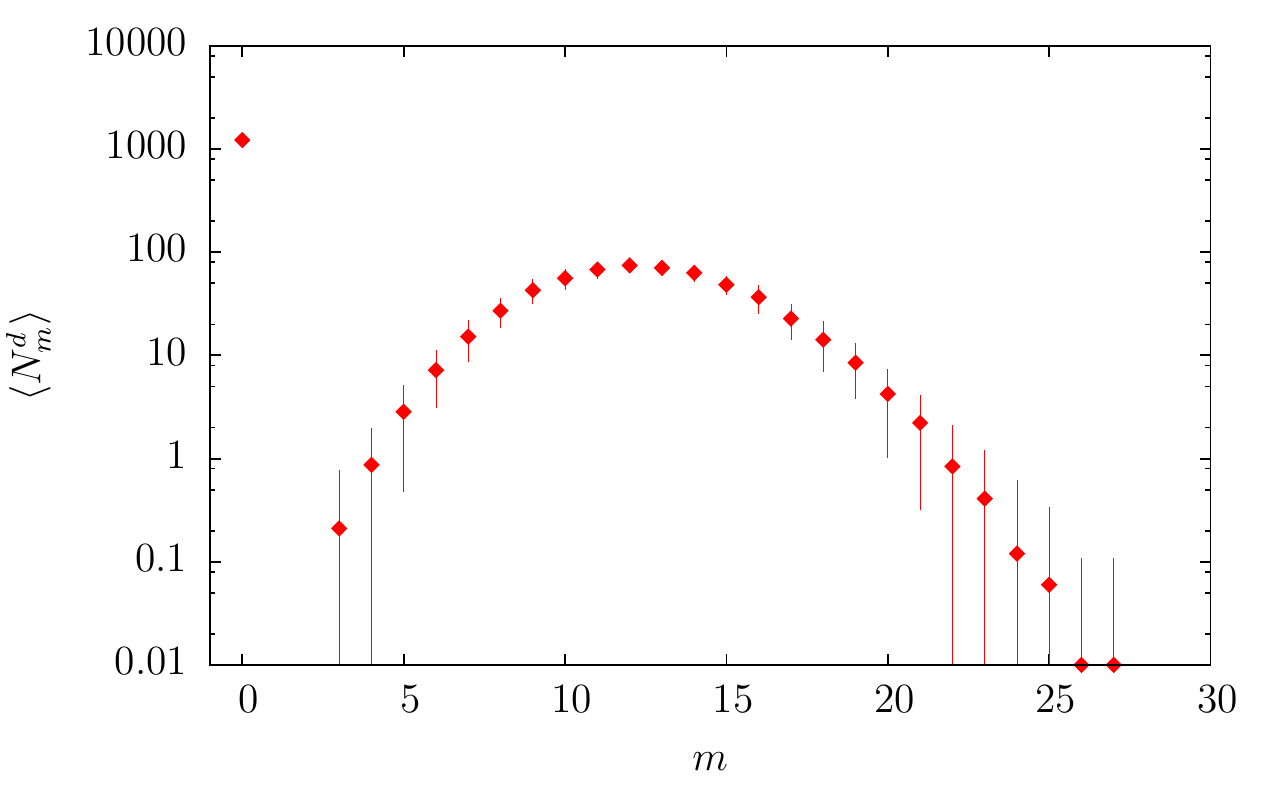}}
\normalsize
\caption{\label{fig:NonManifold} The $\NmCC$ are plotted  for two different  non-manifoldlike causal
  sets. The left hand plot is for a  chain of 10 elements and the right hand plot an average
    over $100$ realisations of an $100$-element KR-order. }
\end{figure}

There are several types of non-manifoldlike causal sets that can be examined in this manner. The
first that come to mind is the totally ordered poset or chain and the totally unordered poset or
antichain. The $\NmCC$  for the former has a simple linearly decreasing behaviour with $m$ as
depicted by the left hand plot of Fig. \ref{fig:NonManifold}, while all the $\NmCC$ for the latter
are simply zero. However, apart from such exotic causal sets, one is interested in what $\NmCC$
looks like for a more typical causal set. Here, we are aided by analytic results which tell us that
as $N$ becomes large, the set of causal sets is dominated by those that are of the
Kleitman-Rothschild or KR form.   A sketch of a small KR- order is shown in Fig. \ref{fig:KR}.
These are distinctly non-manifoldlike since they possess only three ``moments of time''.  
 A typical KR orders has three layers with roughly $N/4$ minimal and maximal elements each and $N/2
 $-elements in the middle layer.   Each minimal and maximal element  are linked to roughly half the
 elements in the middle layer and every minimal element is related to every maximal element. 
Using David Rideout's  Cactus thorn to generate KR orders, we perform  $100$ realisations of
$N=100$ element KR orders to obtain $\langle \NmCC \rangle$, which we  show in Fig. \ref{fig:NonManifold}


\section{\label{sec:conclusion}Conclusions and Outlook}
In this work we demonstrated that interval abundances $\NmCC$ in a causal set $C$ provide an
important class of observables for causal set theory. In particular, by comparing with the
expectation value of the interval abundance $\Nmd$ for an ensemble of causal sets obtained via a
Poisson sprinkling into a flat spacetime interval, we showed that the $\NmCC$ can be used to obtain
``local'' regions or sub-causal sets $C' \subset C$ for a $C$ which faithfully embeds into a general
curved spacetime. Conversely, the existence of local regions in a causal set is a
necessary test for manifoldlikeness of $C$ and as a new continuum dimension estimator.

We began by obtaining closed form expressions for $\Nmd$, and showed that the ratio $\Nmd/\Nzerod$
is independent of the size $N$ of the causal set to leading order. This scale invariance reflects
that of flat spacetime and suggests a rigidity condition encoded by this class of observables.  This
prompted us to conjecture that knowing the $\NmCC$ for $N>>1$ is sufficient to determine if $C$
faithfully embeds into a flat spacetime interval of a given dimension.  We tested these ideas with
extensive simulations.  We found that even for a relatively small ensemble of causal sets obtained
via a Poisson sprinkling into an interval in Minkowski spacetime, the expectation value of the
interval abundances matches very well with our analytic curves for $\Nmd$. In addition the agreement
is very good even for a single causal set, up to the expected Poisson fluctuations in the size of the
causal set, $N \pm \sqrt{N}$, as shown in Fig. (\ref{fig:singlesprinkle}).  This suggests a
prescription for extracting the continuum spacetime dimension from a causal set and thus a necessary
condition for it to be faithfully embeddable into Minkowski spacetime.

In curved spacetime we considered both FRW and deSitter spacetimes. The simulations agree with the
 $\Nmd$ up to fluctuations as long as the scale of flatness is large, but deviate strongly from it when
 the  scale of flatness is small. In the former case we found that the causal set represents a local or
approximately flat spacetime region while in the latter case it is distinctly not local.  We also
examined the effect of topology on the $\Nmd$ and found that again there is agreement with $\Nmd$ only
if the region explored in the spacetime is local.

We then examined a class of causal sets generated via transitive percolation to see if they passed our
test for manifoldlikeness. In \cite{Ahmed:2009qm} it was claimed that a class of such causal sets
possess manifoldlike properties and have a Myrheim-Myer manifold dimension of 3 or 4. We tested 
several of these examples and found that they fail our test since the $\NmCC$ do not agree with the
$\Nmd$ for {\it any} $d$. Hence we concluded that these causal sets are almost certainly {\it not}
manifoldlike. However, it is possible that manifoldlikeness emerges after coarse
graining. Preliminary tests showed  that this is still not the case,  but a more detailed study is
currently underway \cite{percolation}. 

Our simulations thus provide strong support for the rigidity conjecture in Section \ref{sec:test},
namely that knowing $\NmCC$ is sufficient to determine whether $C$ faithfully embeds into a flat
spacetime interval or not.  However, the question arises whether and why this class of observables
is more special than others. For one, it does provide an entire class of observables, and this
itself is useful.  But this is also the case for the abundances of chains $\Cm$, where an $m$-chain
is a totally ordered $m$-element subset of $C$. Thus, a similar analysis may be possible using the
$\Cm$.  However, what distinguishes $m$-intervals from $m$-chains is that the former do in fact
encode Lorentz-invariant {\it local} information while this is not true of the latter.  In
particular, the set of $m$-element intervals with fixed future end point $q$ in a causal set that
faithfully embeds into flat spacetime ``layer'' the past light cone of $q$ along its past invariant
hyperbolae, for each $m$. For example, elements which are linked to $q$ lie within a volume $\sim
\rho^{-1}$ to the past of $q$. Thus, as $m$ increases, one explores regions further and further
from the past null cone boundary of $q$.  However, a chain lacks the same local information. For
example a relation or $2$-chain $p<q$ could either be a link with $|I[p,q]|=0$ or separated by a
very large interval size $|I[p,q]|>>1$.  Thus, the number of relations to the future or the past of
$p$ can lie arbitrarily far from the boundary of the light cones from $p$ -- they are {\it not}
nearest neighbours even in the Lorentzian sense. We believe it is this Lorentz-invariant locality of
the $\NmCC$ which make them useful in defining locality. There are no other obvious candidates for
families of observables and it is therefore tempting to conclude that the $\NmCC$ are unique in this
sense.

It is relatively straightforward to extend these calculations to a region of small curvature using
Riemann normal coordinates and the techniques of \cite{Khetrapal:2012ux,Roy:2012uz}. However, the
expressions for the $\Nmd$ are far more complex, and extracting even an analytic curve from them
requires more computationally intensive tools than in the flat spacetime case. We leave
such investigations to future work where effects of curvature can be studied in greater detail than
in the present work.
  
Finally, local regions in a manifoldlike causal set $C$ could in principle be used to define a
covering $\{ C_i\} $ of $C= \cup_i C_i$, from which a nerve simplicial complex can be
constructed. In the continuum, given a manifold $M$, a nerve simplicial complex can be obtained via a
locally finite convex cover $\mathcal O=\{ O_i\}, M=\cup_iO_i$ , i.e., a cover in which (i)
each set is convex, so that there exists a unique geodesic between any two points in the set (ii)
every $x\in M$ is contained in a finite number of elements of $\mathcal O$.  The nerve simplicial
complex is obtained from $\mathcal O$ by mapping each $k$-wise intersection of sets in $\mathcal O$
to a $k$-simplex.  This simplicial complex is then homotopic to $M$ as shown in
\cite{nerve}. In \cite{homology} a nerve simplicial complex was constructed to obtain the
homology of spatial slices in both the continuum and in a causal set.  One of the main obstructions
to extending it to the full continuum spacetime or the full causal set is that a locally finite cover
built out of Alexandrov intervals does not have an obvious local characterisation. In particular,
there is no way of distinguishing a convex from a non-convex Alexandrov interval purely order
theoretically. Our prescription for locality in the {\it discrete} case however overcomes this
difficulty and it would be interesting to see if the spacetime homology could in fact be recovered
from such a local covering of a  causal set.

Our work opens up several new arenas in the study of discrete causal structure, some of which
may lead us closer to answering  fundamental questions in causal set theory.

\vskip0.5cm
\noindent{{\bf Acknowledgements:}} We thank Rafael Sorkin for discussions and David Rideout for help
with the Cactus code, which we used extensively. The authors would like to thank the ICTP in Trieste for their kind hospitality during parts of this work, L.G. would also like to thank the Raman Research Institute in Bangalore for a very productive visit.
\begin{appendices}

\section{\label{app:closedform}A general formula for hypergeometric functions}
After solving the integration for the interval abundance in Section \ref{sec:flat} it is necessary
to find a closed form expression for sums of the type 
\begin{equation}
	\sum\limits_{n=0}^{\infty} \frac{ (- \rho V)^{n}}{n!}\left( \prod\limits_{i=1}^{q} \frac{1}{n+a_{i}}\right) \frac{\G{x(n+c)}}{\G{x(n+c)+m_{1}}}\frac{\G{x(n+e)}}{\G{x(n+e)+m_{2}}}
\end{equation}
In this expression it is assumed that $m_{1},m_{2}$ are integers.
We can then rewrite it as
\begin{equation}
		\sum\limits_{n=0}^{\infty} \frac{ (- \rho V)^{n}}{n!}\left( \prod\limits_{i=1}^{q} \frac{1}{n+a_{i}}\right) 
		\left( \prod\limits_{k=0}^{m_{1}-1} \frac{1}{x(n+c)+k}\right)\left( \prod\limits_{l=0}^{m_{2}-1} \frac{1}{x(n+e)+l}\right)
\end{equation}
To express this in a closed form we factorise out the $x$ and rewrite the products as gamma
functions which gives 
\begin{align}
		\sum\limits_{n=0}^{\infty} \frac{ (- \rho V)^{n}}{n!} x^{m_{1}+m_{2}} \left( \prod\limits_{i=1}^{q} \frac{\G{n+a_{i}}}{\G{n+a_{i}+1}} \right)  \times \nonumber \\
		\left( \prod\limits_{k=0}^{m_{1}-1} \frac{\G{n+c+\frac{k}{x}}}{\G{n+c+\frac{k}{x}+1}} \right)\left( \prod\limits_{l=0}^{m_{2}-1} \frac{\G{n+e+\frac{l}{x}}}{\G{n+e+\frac{l}{x}+1}}  \right)
\end{align}
This can be rewritten using Pochhammer symbols $\Poch{a}{n}=\G{n+a}/ \G{a}$. Taking all the
factors independent of $n$ out of the sum leads to 
\begin{align}
		x^{m_{1}+m_{2}} \left( \prod\limits_{i=1}^{q} \frac{\G{a_{i}}}{\G{a_{i}+1}} \right) 		\left( \prod\limits_{k=0}^{m_{1}-1} \frac{\G{c+\frac{k}{x}}}{\G{c+\frac{k}{x}+1}} \right)\left( \prod\limits_{l=0}^{m_{2}-1} \frac{\G{e+\frac{l}{x}}}{\G{e+\frac{l}{x}+1}}  \right)  \times \nonumber \\
		\sum\limits_{n=0}^{\infty} \frac{ (- \rho V)^{n}}{n!}\left( \prod\limits_{i=1}^{q}
                  \frac{\Poch{a_{i}}{n}}{\Poch{a_{i}+1}{n}} \right) 		\left(
                  \prod\limits_{k=0}^{m_{1}-1}
                  \frac{\Poch{c+\frac{k}{x}}{n}}{\Poch{c+\frac{k}{x}+1}{n}} \right)\left(
                  \prod\limits_{l=0}^{m_{2}-1}
                  \frac{\Poch{e+\frac{l}{x}}{n}}{\Poch{e+\frac{l}{x}+1}{n}}  \right), 
\end{align}
which can then be reexpressed in terms of the  $_{l+m_{1}+m_{2}}F_{l+m_{1}+m_{2}}$ hypergeometric
function 
\begin{align}
		\left( \prod\limits_{i=1}^{q} \frac{1}{a_{i}} \right) 		\left( \prod\limits_{k=0}^{m_{1}-1} \frac{1}{x c+k} \right)\left( \prod\limits_{l=0}^{m_{2}-1} \frac{1}{x e+l}  \right) \times 	\nonumber \\
		\mFm{l+m_{1}+m_{2}}{[a_{i}],[c+\frac{k}{x}],
                  [e+\frac{l}{x}]}{[a_{i}+1],[c+\frac{k}{x}+1],[ e+\frac{l}{x}+1]}{-\rho V}, 
\end{align}
where the $[a_{i}]$ stand for $a_{1},\ldots,  a_{q}$ and $[c+\frac{k}{x}]$ (respective $[e+\frac{l}{x}]$) do stand for $m_{1}$ ($m_{2}$) terms in which $k$ ($l$) varies from $0$ to $m_{1}-1$ ( $0$ to $m_{2}-1$).
One last simplification allows us to write
\begin{align}
	\left(\prod\limits_{i=1}^{q} \frac{1}{a_{i}} \right)  \quad \frac{\G{x c} \G{x e}}{\G{x c +m_{1}} \G{x e +m_{2}}}
	\times 	\nonumber \\
	\mFm{l+m_{1}+m_{2}}{[a_{i}],[c+\frac{k}{x}], [e+\frac{l}{x}]}{[a_{i}+1],[c+\frac{k}{x}+1],[ e+\frac{l}{x}+1]}{-\rho V}
\end{align}
If $c, e, a_{i}$ have an integer distance smaller than $m_{1}$ ($m_{2}$) it is possible to simplify
this further since arguments of the hypergeometric function that arise on both sides cancel each
other. In our calculation these simplifications will indeed take place but details are specific to
each case.

\section{\label{app:diff} Deriving the $m \neq 0$ case from the $m=0$ case using Hypergeometric function identities}
We can derive $\Nmd$ from $\Nzerod$ by taking derivatives.
The expression for $N^{d}_{0}$ is of the form
\begin{equation} \label{eq:solForm}
\chi  \;(\rho V)^{2} \mFm{p}{a_{1},\dots,a_{p}}{b_{1},\dots,b_{p}}{-\rho V} \;, 
\end{equation}
where  we have lumped some of the dimension dependent constants into the term $\chi$.
Using \eqref{eq:genm}  the expression for the $\Nmd$ is  
\begin{equation} \label{eq:solForm}
 \chi \frac{\left( -\rho\right)^{m+2}}{m!} \pd{^{m}}{\rho^{m}}  \; V^{2} \mFm{p}{a_{1},\dots,a_{p}}{b_{1},\dots,b_{p}}{-\rho V}\;.
\end{equation}
We use the identity \cite{wolfram} 
\begin{equation}
	\pd{^{m}}{z^{m}}\mFm{p}{a_1,\dots,a_p}{b_1,\dots,b_p}{z}=\frac{\prod _{j=1}^p \Poch{a_j}{m} }{\prod _{j=1}^p \Poch{b_j}{m}} \mFm{p}{m+a_1,\dots,m+a_p}{m+b_1,\dots,m+b_q}{z} \;,
\end{equation}
to simplify Eqn. (\ref{eq:solForm}) to 
\begin{equation}
	\chi \frac{ (\rho V)^{m+2}}{m!} \frac{\prod _{j=1}^p \Poch{a_j}{m} }{\prod _{j=1}^p \Poch{b_j}{m}}  \mFm{p}{a_{1}+m,\dots,a_{p}+m}{b_{1}+m,\dots,b_{p}+m}{-\rho V}
\end{equation}
This expression allows for further simplifications, depending on the $a_{j}, b_{j}$. These can be done for each individual case.

\section{\label{app:continuum} Derivation of the Continuum Limit}

The quantity we calculate is
\begin{equation} \label{eq:limitdef}
S_m^d\equiv \lim_{\rho \rightarrow \infty} \frac{\Nmd(\rho, V)}{\Nzerod(\rho, V)}\; , 
\end{equation}
where $\Nmd$  is given by Eqn. (\ref{eq:flatclosedform}). 
To investigate the $N\rightarrow \infty$ limit of the $\Nmd$ we need a large $N=\rho V$ expansion of
the hypergeometric functions that appear in Eqn. (\ref{eq:flatclosedform}), which when appropriately
rearranged are of the form
\begin{align}\label{eq:shape}
\mFm{d}{a_1, \dots, a_d }{a_1+2,\dots , a_d+2}{-N} \;, \quad a_i=\frac{2i}{d} + m, \, \,i=1, \ldots d-1,
\quad  a_d=1+m.
\end{align} 
We make repeated use of the identity Eqn. (\ref{eq:pFq}) as well as the identity Eqn. (\ref{eq:1F1}). 
Since the first identity can not be used if two of the $a_i$ are equal, or equal up to an integer
whose absolute value is smaller than $m_i$,  we will need to be careful in even dimensions. For
$d=2$, in particular  we need a different approach. We will thus treat odd and even dimensions separately and $d=2$ as a separate case.

\subsection{Odd dimensions:}
This involves the most straightforward application of Eqn. \eqref{eq:pFq}  to expand
\eqref{eq:shape}: 
\begin{align}
&\,\mFm{d}{a_1, \dots, a_d }{a_1+2,\dots , a_d+2}{-z} = \prod _{j=1}^d \Poch{a_j}{2} \sum _{k=1}^d
\sum _{j_1=0}^{1} \ldots  \sum _{j_d=0}^{1} \frac{1}{a_k+j_k} \nno \\
&\prod _{l=1}^d \left(-1\right)^{j_l} \prod _{\substack{i=1\\ i\neq k}}^d \frac{1}{a_i+j_i-a_k-j_k} \, \, \, \pFq{1}{1}{a_k+j_k}{a_k+j_k+1}{-z} \label{eq:step1}
\end{align}

For us $m_l =2 \; j_l=0,1$ so that  $\Poch{1-m_l}{j_l}=(-1)^{j_l}$.  Using Eqn. \eqref{eq:1F1} in \eqref{eq:step1}. 
\begin{align}
=& \sum _{k=1}^d \sum _{j_1=0}^{1} \ldots  \sum _{j_d=0}^{1}  \, (z)^{-a_k-j_k}
\left(\G{a_k+j_k}-\G{a_k+j_k,z}\right) \prod _{l=1}^d \Poch{a_l}{2} \left(-1\right)^{j_l} \prod
_{\substack{i=1\\ i\neq k}}^d \frac{1}{a_i+j_i-a_k-j_k}\;. 
\label{eq:Junkone} 
\end{align}
For $z \to \infty$ the terms containing $\G{a,z}$ fall off like $e^{-z} z^{a-1}$,
c.f. Eqn. \eqref{eq:Gaz} and do  not contribute in the large $z$ limit.  Thus, to leading order the  hypergeometric function is a power series with terms $z^{-a_i-j_i}$.
The leading order term is therefore $a_1=\frac{2}{d}+m$, $j_1=0$, while  the next to leading order
is $a_d=1+m$ for $d=3$ and $a_2=\frac{4}{d}+m$ for $d \geq 5$.
We then only need to calculate the case $k=1, j_{k}=0$.

Combining the products in Eqn. (\ref{eq:Junkone}), we then sum  over the $j_i$,
\begin{align}
&\sum_{j_i=0}^1 \frac{\Poch{a_i}{2} (-1)^{j_i} }{a_i+j_i-a_1}=\Poch{a_i}{2}\left(\frac{1}{a_i-a_1}-\frac{1}{a_i+1-a_1}\right)=\frac{ \Poch{a_i}{2} }{(a_i-a_1)(a_i+1-a_1)} 
\end{align}
 after which we take the product over $i$ to obtain 
\begin{align}
	&\Poch{a_k}{2} \prod_{\substack{i=1\\ i\neq k}}^d \frac{ \Poch{a_i}{2} }{(a_i-a_k)(a_i+1-a_k)} =  \Poch{\frac{2}{d}+m}{2} {\frac{ \Poch{m+1}{2} }{(1-\frac{2}{d})(2-\frac{2}{d})} }\prod_{i=2}^{d-1} \frac{\Poch{\frac{2}{d} i+1}{2}}{\frac{2}{d}(i-1)(\frac{2}{d}(i-1)+1)}
	\intertext{ We do the product for the different parts separately:}
 & \Poch{\frac{2}{d}+m}{2} \prod_{i=2}^{d-1} \Poch{\frac{2}{d} i+1}{2 } = \prod_{i=1}^{d-1} \left(
   \frac{2}{d}i+m \right) \left( \frac{2}{d}i+m+1 \right) \nno \\
&= \left(\frac{2}{d}\right)^{2d-2} \Poch{\frac{d}{2}m+1}{d-1} \Poch{\frac{d}{2}(m+1)+1}{d-1} \nno\\
&\prod_{i=2}^{d-1}\frac{1}{\frac{2}{d}(i-1)+1} \frac{d}{2(i-1)}= \left(\frac{2}{d}\right)^{-2d+2}  \frac{1}{\G{d-1} \Poch{\frac{d}{2}-1}{d-2}}
\end{align}
These are  combined to find 
\begin{align}
z^{-\frac{2}{d}-m} \frac{  2 \G{\frac{2}{d}+m} (m+1)(m+2) }{\G{d} (d-2)) \Poch{\frac{d}{2}+1}{d-2}}
\Poch{\frac{d}{2}(m+1) +1}{d-1} \Poch{\frac{d}{2}m+1}{d-1} + \begin{cases} \ \ \
  \mathcal{O}(z^{-m-1}) & \mbox{if } d=3 \\ \ \ \ \mathcal{O}( z^{-\frac{4}{d}-m} ) & \mbox{if } d
  \geq 5 \end{cases} 
\end{align}
Inserting this into Eqn. \eqref{eq:flatclosedform} for large $N $ gives  
\begin{align}
N_{m}^{d}(N)=& \frac{N^{2-\frac{2}{d}}}{m!} \G{\frac{2}{d}+m} \frac{\G{d} }{ \left(\frac{d}{2}
    -1\right) \Poch{\frac{d}{2}+1}{d-2}} + \begin{cases} \ \ \ \mathcal{O}(N) & \mbox{if } d=3 \\ \
  \ \ \mathcal{O}( N^{2-\frac{4}{d}} ) & \mbox{if } d \geq 5 \end{cases},  
\end{align}
which gives us Eqn. (\ref{eq:limitdeftwo}). 

\subsection{Even dimensions:}
In even dimensions it is possible for   two of the $a_i$ to be  equal, or equal up to an integer
whose absolute value is less than $m_i$. This therefore requires more care.
The arguments of the hypergeometric functions in Eqn. (\ref{eq:flatclosedform}) however do admit a
non-degenerate split: 
\begin{equation}
\mFm{d}{\frac{2}{d}+m, \cdots , 1-\frac{2}{d}+m, m+1, \frac{2}{d}+m+1, \cdots , 2-\frac{2}{d}+m,
  m+1}{\underbrace{\frac{2}{d}+m+2, \cdots , 3-\frac{2}{d}+m}_{\text{ the first $\frac{d}{2}-1$
      terms}},\underbrace{ m+3}_{\text{the $\frac{d}{2}$th term}}, \underbrace{\frac{2}{d}+m+3,
    \cdots , 4-\frac{2}{d}+m}_{\text{ the $\frac{d}{2}-1$ terms before the last}}, m+3}{-z}, 
\end{equation}
which can be shuffled to simplify the calculation. Exchanging the first $\frac{d}{2}-1$ terms and
the $\frac{d}{2}-1$ terms before the last in the upper row, changes  the relationship between the top and
bottom row.
Instead of a hypergeometric function of the form \eqref{eq:shape} we now have one of the form
\begin{equation}
\mFm{d}{a_1 ,\cdots, a_{\frac{d}{2}-1},a_{\frac{d}{2}},a_{\frac{d}{2}+1},\cdots ,a_{d-1},a_d}{a_1+1 ,\cdots, a_{\frac{d}{2}-1}+1,a_{\frac{d}{2}}+2,a_{\frac{d}{2}+1}+3,\cdots ,a_{d-1}+3,a_d+2}{-z}
\end{equation}
We now proceed in two steps.
The first is to use \eqref{eq:pFq} on the first $\frac{d}{2}-1$ terms. Here $n=d/2-1$, $m_i=1
\forall i \in [1, \ldots, d/2-1]$ and hence $j_i=0 \forall i$.  Thus    
\begin{align}\label{eq:alphaexp}
&\mFm{d}
{\frac{2}{d}+m+1, \cdots , 2-\frac{2}{d}+m, m+1, \frac{2}{d}+m, \cdots , 1-\frac{2}{d}+m, m+1}
{\frac{2}{d}+m+2, \cdots , 3-\frac{2}{d}+m,m+3, \frac{2}{d}+m+3, \cdots , 4-\frac{2}{d}+m, m+3}{-z}
= \nno \\
&\suml{\alpha = 1}{\frac{d}{2}-1} \mFm{\frac{d}{2}+2}
{\frac{2}{d} \alpha +m+1,  \frac{2}{d}+m, \cdots , 1-\frac{2}{d}+m,m+1, m+1}
{\frac{2}{d} \alpha +m+2,\frac{2}{d}+m+3, \cdots , 4-\frac{2}{d}+m,m+3,  m+3}{-z} 
 \prod\limits_{\substack{j=1 \\ j\neq \alpha} }^{\frac{d}{2}-1} \frac{j+\frac{d}{2}(m+1)}{j-\alpha} 
\end{align}
Next we apply eqn. \eqref{eq:pFq} for a second time, with $i=2, \ldots d/2+1$. Now $n=d/2$ and
$m_i=3 $ for $i \in [2, \ldots, d/2]$ while $m_{d/2+1}=2$. Thus, 
\begin{align}
&\mFm{\frac{d}{2}+2}
{\frac{2}{d} \alpha +m+1,  \frac{2}{d}+m, \cdots , 1-\frac{2}{d}+m,m+1, m+1}
{\frac{2}{d} \alpha +m+2,\frac{2}{d}+m+3, \cdots , 4-\frac{2}{d}+m,m+3,  m+3}{-z} =\nno \\
&\left(\prod_{j=1}^{\frac{d}{2}} \frac{\Poch{\frac{2}{d}j+m}{3}}{2} \right) \cdot \frac{2}{m+3} \sum_{k=1}^{\frac{d}{2}}
\sum_{j_{1}}^{2}\cdots\sum_{j_{\frac{d}{2}-1}}^{2}\sum_{j_{\frac{d}{2}}}^{1} \frac{(-1)^{j_{\frac{d}{2}}}}{\frac{2}{d} k+j_{k}+m} 
\left( \prod_{l=1}^{\frac{d}{2}-1} \frac{ \Poch{-2}{j_l}}{j_l!} \right) \times \nno \\
& \left( \prod_{\substack{i=1\\ i\neq k}}^{\frac{d}{2}} \frac{1}{\frac{2}{d}(i-k)+j_i-j_k} \right) 
\mFm{3}{\frac{2}{d}k+m+j_k, \frac{2}{d}\alpha+m+1, m+1}{\frac{2}{d}k+m+j_k+1, \frac{2}{d}\alpha+m+2, m+3}{-z}
\end{align}
To take the limit $z\to \infty$ we need to expand this $_3 F_3$ for large $z$. We cannot do the entire expansion because of the special cases where $\frac{2}{d}k+m+j_k =\frac{2}{d}\alpha +m+1$ and $\frac{2}{d}k+m+j_k = m+1$. We thus need to make the expansion for three different possibilities:
\begin{itemize}
\item $\frac{2}{d}k+m+j_k \neq \frac{2}{d}\alpha +m+1 \neq m+1$ in this case we can find an exact expansion for all $z$
\item $\frac{2}{d}k+m+j_k =\frac{2}{d}\alpha +m+1 $ for this case we can take an expansion in large $z$ and find that it's leading order contributions in that limit are $z^{-1-m}\log{z}$.
\item $\frac{2}{d}k+m+j_k = m+1$ we can again take a large $z$ expansion and find terms proportional to $z^{-m-1}\log{z}$ in leading order.
\end{itemize}

The expansions for the first case, $\frac{2}{d}k+m+j_k \neq \frac{2}{d}\alpha +m+1 \neq m+1$, can be found using Eqn. \eqref{eq:pFq}
\begin{align}\label{eq:firstcase}
&\mFm{3}{\frac{2}{d}k+m+j_k, \frac{2}{d}\alpha+m+1, m+1}{\frac{2}{d}k+m+j_k+1,
  \frac{2}{d}\alpha+m+2, m+3}{-z}= \nno \\
& e^{-z} \frac{(2+m)(\frac{2}{d}\alpha+m+1)(\frac{2}{d}k+m+j_k)}{\frac{2}{d} \alpha (\frac{2}{d} k+j_k-1)} \nno \\
& - \frac{(1+m)(2+m) (\frac{2}{d}\alpha+m+1)(\frac{2}{d}k+m+j_k)}{(\frac{2}{d}\alpha-1)(\frac{2}{d}k+j_k-2)} z^{-2-m} \left( \G{2+m}-\G{2+m,z} \right) \nno \\
& + \frac{(2+m) (\frac{2}{d}\alpha +m+1) (\frac{2}{d}k+m+j_k)}{\frac{2}{d}\alpha (\frac{2}{d}k+j_k-1)} z^{-1-m} \left( \G{2+m}-\G{2+m,z}\right) \nno \\
&- \frac{ (2+m)(1+m)(\frac{2}{d} \alpha+m+1) (\frac{2}{d}k+m+j_k)}{\frac{2}{d}\alpha (\frac{2}{d}\alpha-1)(\frac{2}{d}(\alpha-k)+1-j_k)}z^{-1-m-\frac{2}{d}\alpha} \left( \G{\frac{2}{d}\alpha +1+m} - \G{\frac{2}{d}\alpha+1+m,z} \right)	\nno \\
&-\frac{(2+m)(1+m) (\frac{2}{d}\alpha +m+1)(\frac{2}{d}k+m+j_k)}{(\frac{2}{d}k+j_k-2)(\frac{2}{d}k+j_k-1)(\frac{2}{d}(k-\alpha)+j_k-1)}z^{-j_k-m-\frac{2}{d}k} \left(\G{j_k+m+\frac{2}{d}k}-\G{j_k+m+\frac{2}{d}k,z}\right) 
\end{align}
For the second and third case we use Mathematica to obtain the expansion.
In the second case,$\frac{2}{d}k+m+j_k =\frac{2}{d}\alpha +m+1 $ ,
\begin{align}\label{eq:secondcase}
&\mFm{3}{\frac{2}{d}\alpha +m+1, \frac{2}{d}\alpha+m+1, m+1}{\frac{2}{d}\alpha +m+2, \frac{2}{d}\alpha+m+2, m+3}{-z}=\nno \\
&\frac{(m+2) (\frac{2}{d}\alpha +m+1)^2 }{(\frac{2}{d}\alpha )^2} z^{-m-1} \Gamma (m+2)
-\frac{(m+1) (\frac{2}{d}\alpha +m+1)^2}{(\frac{2}{d}\alpha -1)^2} z^{-m-2} \Gamma (m+3)\nno \\
&+\frac{(m+1) (m+2) (\frac{2}{d}\alpha +m+1)}{(\frac{2}{d}\alpha -1) \frac{2}{d}\alpha } \Gamma (m+\frac{2}{d}\alpha +2) z^{-\frac{2}{d}\alpha -m-2} (\frac{2 \frac{2}{d}\alpha -1}{(\frac{2}{d}\alpha -1) \frac{2}{d}\alpha }+ \psi ^{(0)}(m+\frac{2}{d}\alpha +1)+ \log (z)) \nno \\
& +\cdots
\end{align}
and in the third case, $\frac{2}{d}k+m+j_k =m+1 $ ,
\begin{align}\label{eq:thirdcase}
&\mFm{3}{m+1, \frac{2}{d}\alpha+m+1, m+1}{m+2, \frac{2}{d}\alpha+m+2, m+3}{-z}=\nno \\
& - \frac{(1+m)^2 (2+m)(1+m+\frac{2}{d}\alpha)}{(\frac{2}{d}\alpha)^2(\frac{2}{d}\alpha-1)} z^{-1-m-\frac{2}{d}\alpha} \G{1+m+\frac{2}{d} \alpha} \nno \\
&-\frac{(m+1)(1+m+\frac{2}{d}\alpha)}{(\frac{2}{d}\alpha)^2} z^{-1-m} \G{3+m}\left( 2-\log{z} +\Psi^{(0)}(m+1) \right)+ \cdots
\end{align}

The leading order terms are $z^{-\frac{2}{d}-m}$, for $k=1, j_k=0$ from
Eqn. (\ref{eq:firstcase}). The next to leading order term is  $z^{-\frac{4}{d}-m}$ if $d>4$, while
for $d=4$ it is $z^{-m-1} \log{z}$,  from \eqref{eq:secondcase} and \eqref{eq:thirdcase}.
Thus to  leading order we find
\begin{align} \label{eq:insertexp}
&\left(\prod_{j=1}^{\frac{d}{2}} \frac{\Poch{\frac{2}{d}j+m}{3}}{2} \right) \cdot \frac{2}{m+3}
\sum_{j_{1}}^{2}\cdots\sum_{j_{\frac{d}{2}-1}}^{2}\sum_{j_{\frac{d}{2}}}^{1} \frac{1}{\frac{2}{d} 1+m} 
\left( \prod_{l=1}^{\frac{d}{2}-1} \frac{ \Poch{-2}{j_l}}{j_l!} \prod_{i=2}^{\frac{d}{2}-1} \frac{1}{\frac{2}{d}(i-1)+j_i} \right) \times\nno \\ 
& \frac{(-1)^{j_{\frac{d}{2}}} }{1+j_{\frac{d}{2}}-\frac{2}{d}}
\frac{(2+m)(1+m) (\frac{2}{d}\alpha +m+1)(\frac{2}{d}+m)}{(\frac{2}{d}-2)(\frac{2}{d}-1)(\frac{2}{d}(\alpha-1)+1)}z^{-m-\frac{2}{d}} \G{m+\frac{2}{d}} 
\end{align}

The products in \eqref{eq:insertexp} simplify as 
\begin{align}\label{eq:sums}
&\prod_{j=1}^{\frac{d}{2}} \frac{\Poch{\frac{2}{d}j+m}{3}}{2}= 2^d d^{-\frac{3}{2}d} \frac{\G{\frac{d}{2}(m+3)+1}}{\G{\frac{d}{2}m+1}} \nno \\
&\sum_{j_l=0}^2 \frac{\Poch{-2}{j_l}}{2 j_l!(\frac{2}{d}(l-1)+j_l)}= \frac{d^3}{4 (l-1) (l-1+d)
  ((l-1)+\frac{d}{2} )}\nno \\
&\prod_{l=2}^{\frac{d}{2}-1}\frac{d^3}{4 (l-1) (l-1+d) ((l-1)+\frac{d}{2} )}= \frac{ 2^{2-d} (d-2)(d-1) d^{-4+\frac{3}{2}d}}{\G{\frac{3}{2}d-1}}
\end{align}

so that \eqref{eq:insertexp} simplifies to 
\begin{equation} \label{eq:sweet60}
\frac{d \;\G{\frac{d}{2}(m+3)}}{(d-2)(d-1)} \frac{\G{\frac{2}{d}+m} (m+1)(m+2)}{\G{\frac{d}{2} m+1} \G{\frac{3}{2}d-1}} 
\frac{ \frac{d}{2}(m+1) +\alpha}{\frac{d}{2}+\alpha -1} z^{-\frac{2}{d}-m} .
\end{equation}
Inserting this into Eqn. \eqref{eq:alphaexp} we can perform the summation over $\alpha$: 
\begin{align}
&\suml{\alpha = 1}{\frac{d}{2}-1} \frac{ \frac{d}{2}(m+1) +\alpha}{\frac{d}{2}+\alpha -1} 
 \prod\limits_{\substack{j=1 \\ j\neq \alpha} }^{\frac{d}{2}-1} \frac{j+\frac{d}{2}(m+1)}{j-\alpha} 
\end{align}
The product in the above expression gives 
\begin{equation}
(\frac{d}{2}(m+1) +\alpha) \prod\limits_{\substack{j=1 \\ j\neq \alpha} }^{\frac{d}{2}-1} \frac{j+\frac{d}{2}(m+1)}{j-\alpha} =\frac{ (-1)^{\alpha-1} \G{\frac{d}{2}(m+2)}}{\G{\alpha} \G{\frac{d}{2}-\alpha} \G{\frac{d}{2}(m+1)+1}}
\end{equation}
so that the sum reduces to 
\begin{equation}
\suml{\alpha = 1}{\frac{d}{2}-1} \frac{ (-1)^{\alpha-1} \G{\frac{d}{2}(m+2)}}{\G{\alpha} \G{\frac{d}{2}-\alpha} \G{\frac{d}{2}(m+1)+1}} \frac{1}{\alpha -1+\frac{d}{2}} = \frac{\G{\frac{d}{2}} (d-1) \G{\frac{d}{2}(m+2)}}{\G{d} \G{\frac{d}{2}(m+1)+1}} \;.
\end{equation}
Combining this with Eqn. \eqref{eq:sweet60} gives 
\begin{align}
z^{-m-\frac{2}{d}} \frac{d \; \G{\frac{d}{2}}}{(d-2) \G{d} \G{\frac{3}{2}d -1}} \G{\frac{2}{d}+m} (m+1)(m+2) \Poch{\frac{d}{2}(m+1)+1}{d-1}
\Poch{\frac{d}{2}m+1}{d-1} \;.
\end{align}
Thus, we find 
\begin{equation}
 \Nmd(N) =\frac{ N^{2-\frac{2}{d}}}{m!} \frac{\G{d+1}\G{\frac{d}{2}}}{(d-2)\G{\frac{3}{2}d -1}} \G{\frac{2}{d}+m}+ \begin{cases}
 \mathcal{O}(N \log{N}) & \text{ for } d=4\nno \\
 \mathcal{O}(N^{2-\frac{4}{d}}) &\text{ for } d>4 
 \end{cases}
\end{equation}
Realising that
\begin{equation}
\frac{ d \; \G{\frac{d}{2}}}{\G{\frac{3}{2}d-1}} =\frac{2}{\Poch{\frac{d}{2}+1}{d-2}} \;,
\end{equation}
we note that this agrees with the expression for odd dimensions, and we can thus write
\begin{equation}
 \Nmd(N) =\frac{N^{2-\frac{2}{d}}}{m!} \G{\frac{2}{d}+m} \frac{\G{d} }{ \left(\frac{d}{2} -1\right) \Poch{\frac{d}{2}+1}{d-2}}+ \begin{cases}
 \mathcal{O}(N ) & \text{ for } d=3\nno \\
 \mathcal{O}(N \log{N}) & \text{ for } d=4\nno \\
 \mathcal{O}(N^{2-\frac{4}{d}}) &\text{ for } d>4 
 \end{cases}
\end{equation}
for all $d>2$, which gives Eqn. (\ref{eq:limitdeftwo}). 

\subsection{The case $d=2$}
 $d=2$ is a special case for which $_2 F_2$ can be expanded for $z \to \infty$ using Mathematica,
\begin{equation}
\mFm{2}{m+1,m+1}{m+3,m+3}{-z} = z^{-1-m} \G{3+m} (m+1)(m+2) \log{z} + \mathcal{O}(z^{-1-m} ) \;.
\end{equation}
Inserting this in \eqref{eq:flatclosedform} leads to
\begin{equation}
\Nmdv{m}{2}(N)= N \log{N} + \mathcal{O}(N) \;,
\end{equation}

which gives
\begin{equation} \label{eq:limitdef}
\lim_{\rho \rightarrow \infty} \frac{\Nmdv{m}{2}(\rho, V)}{\Nmdv{0}{2}(\rho, V)} = 1 
\end{equation}
and hence to leading order agrees with the expression for $d>2$, so that we have finally recovered
Eqn. \eqref{eq:limitdeftwo} for all $d\geq 2$

\end{appendices}

\bibliographystyle{plain}
\bibliography{abundance}
\end{document}